\documentclass[12pt]{JHEP3}
\usepackage{amsfonts,amssymb}
\usepackage{amsmath}

\usepackage{epsf, graphicx}

\newcommand{\diag}{\hbox{diag}}

\def\T11{{T}^{1,1}}
\def\bear{\begin{eqnarray}}
\def\eear{\end{eqnarray}}

\def\dim{\mathrm{dim}}

\newcommand{\pa}{\partial}
\newcommand{\tr}{{\rm tr}}
\newcommand{\comment}[1]{}
\newcommand{\pasl}{\pa\kern-.55em /}

\newcommand{\ksl}{k\kern-.55em /}

\newcommand{\ket}[1]{|#1\rangle}

\newcommand{\vev}[1]{\langle #1\rangle}

\DeclareFixedFont{\xiiss}{OT1}{cmss}{m}{n}{12}
\DeclareFixedFont{\ixss}{OT1}{cmss}{m}{n}{9}
\DeclareFixedFont{\cmrnine}{OT1}{cmr}{m}{n}{9}
\newcommand{\field}[1]{\mathbb{#1}}

\newcommand{\BC}{{\field C}}
\newcommand{\BR}{{\field R}}
\newcommand{\BZ}{{\field Z}}

\newcommand{\CCs}{\hbox{\ixss C\kern-.4emI}}
\newcommand{\ZZs}{\hbox{\ixss Z\kern-.4emZ}}

\newcommand{\CV}{{\cal V}}

\newcommand{\CP}{{\BC\field P}}

\newcommand{\mU}{{\mathfrak U}}
\newcommand{\mV}{{\mathfrak V}}
\newcommand{\mW}{{\mathfrak W}}
\newcommand{\mZ}{{\mathfrak Z}}

\newcommand{\myfig}[3]{\begin{figure}[ht]
\begin{center}
\leavevmode
\epsfxsize=#2cm
\epsfbox{#1}
\end{center}
\caption{#3}
\label{fig:#1}
\end{figure}}

\title{ Aspects of ABJM orbifolds}
\author{David Berenstein, Mauricio Romo \\
 Department of Physics, UCSB, Santa Barbara, CA 93106}

\keywords{ AdS/CFT, Monopoles}
\abstract{
We study abelian and non-abelian orbifolds of the ABJM model. We compute the precise moduli space of these models by analyzing the classical BPS equations for the theory on the cylinder, which include classical solutions of magnetic monopole operators. These determine the chiral ring of the theory, and thus they provide the complete set of order parameters determining the classical vacua of the theory. We show that
the proper quantization of these semiclassical solutions gives us the topology of moduli space, including the additional quotient information due to the Chern-Simons levels. In general, we find that in the dual M-theory setup, the M-theory fiber is divided by the product of the Chern-Simons level times the order of the orbifold group, even in the non-abelian case. This depends non-trivially on how the different Chern-Simons terms have different levels in these constructions. We also see a direct relation in this setup between the Chern-Simons levels of the different groups and fluxes for fractional brane cycles. We also show that the problem of the moduli space can be much more easily analyzed by using the method of images and representation theory of crossed product algebras rather than dealing only with the quiver theory data.
}

\begin{document}

\section{Introduction}

In the past year, the AdS/CFT correspondence has found a new set of examples in three
dimensional superconformal field theories that share many characteristics with the original
${\cal N}=4$ SYM and its $AdS_5\times S^5$ dual \cite{M} and their orbifolds \cite{KS}.

These theories, whose first example was constructed in \cite{ABJM} and which we will call the ABJM model, have the following properties that make them similar to their four dimensional cousins:
\begin{enumerate}
\item The theories posses an ${\cal N}=2$ supersymmetry in three dimensions: these have a simple superspace description similar to the ${\cal N}=1$ superspace in four dimensions.
\item The degrees of freedom are vector (super)fields and chiral superfields. The gauge groups can be arbitrarily large (they have a rank $N$ that can be taken to be large).
\item The theories have simple lagrangians that are classically conformally invariant with canonical kinetic terms for matter. The vector field lagrangian is of Chern-Simons type. This was originally suggested by Schwarz \cite{Sch} as a source of interesting dualities, but the examples with duals were found later in \cite{ABJM}. The level $k$ plays a similar role to the Yang Mills coupling constant $g_{YM}^{-2}$.
\item The theories admit a large $N$ t' Hooft limit by taking $N\to \infty$ and keeping $\lambda = N/k$ fixed.
\item For small $\lambda$ the theories can be analyzed using standard perturbation theory.
\item For large and finite $\lambda$ the theory can be better thought of as a type IIA string theory in an $AdS_4\times X_6$ geometry.
\item If $k$ is kept fixed and $N$ made very large, the theory can be best described by an M-theory setup on $AdS_4\times X_7$. The seven dimensional space is a circle bundle over $X_6$ that is determined by the level $k$.
\end{enumerate}

Because of these similarities to four dimensional examples, a lot of work has been done at the level of comparisons between both sides of the correspondence following the familiar ideas used in four dimensions. These comparisons usually deal primarily with the $AdS_4\times X_6$ string limit where one usually can calculate the dimensions of operators using perturbation theory.

Unlike their four dimensional cousins, the perturbative gauge invariant elements of the chiral ring are not sufficient to describe the moduli space of vacua. These moduli spaces are essentially $N$ particles on a real cone over $X_7$ as expected by the M-theory setup.
In contrast, perturbative gauge invariant words would give holomorphic spaces of lower dimension than the cone over $X_7$ would demand. In essence, the naive chiral ring made up of gauge invariant polynomials in the holomorphic fields is identical to that of a four dimensional theory. These usually can only describe multiple branes on a Calabi-Yau threefold \footnote{This can be understood in terms of a simple condition: that the F-terms equations  are naturally dual to the superfields  of the theory. When translated into a mathematical framework,  one builds an  associated algebra of a quiver theory: a quiver path algebra with relations. This condition on the F-terms becomes a homological algebra relation that identifies the $Ext^2(A,B)$ functor (relations obtained from F-terms) as a natural dual to $Ext^1(B,A)$ (chiral fields) on modules  \cite{BD}. This is identical to  what one would obtain from Serre duality for threefolds with a trivial canonical bundle.}.

The solution to this puzzle lies in the fact that in three dimensional theories there are additional non-perturbative elements of the chiral ring. These chiral ring operators create magnetic fluxes and have a similar profile to the spatial components of magnetic monopoles in four dimensions. These operators are called magnetic monopole operators. Their presence is necessary to match the spectrum of protected operators of eleven dimensional supergravity on $X_7$ \cite{ABJM}. These carry the quantum numbers of angular momentum on the circle fiber of $X_7$ over $X_6$. From the point of view of IIA string theory seen as a compactification of M-theory on a circle, the dual particles to these operators describe D0-branes in the bulk and not strings, as the simplest gauge invariant observables do.

From the point of view of calculating the moduli space of vacua from the field theory, the vacuum expectation values of these non-perturbative operators are some of the order parameters distinguishing the different points in the moduli space. This means that non-perturbative effects are crucial to the understanding of the model, even at the level of describing the precise shape of the moduli space of vacua. This is very unlike the examples in four dimensions, where knowledge of the perturbative spectrum is enough to describe the moduli space.
Very importantly, the topology of the moduli space of vacua depends on $k$. This is another way to understand why the level of Chern -Simons terms in the lagrangian should be quantized.

The difference in the dimension of the moduli space from what one can guess perturbatively can be qualitatively explained by stating that the electric-magnetic dual of a vector super-particle in three dimensions is a complex scalar. It is the vacuum expectation values of these dual scalars that one needs to probe to completely characterize this moduli space. This is why we require understanding magnetic monopole instanton effects or operators in three dimensions to fully address this issue: they have to be the non-perturbative probes that can probe the field of a dual electromagnetic field. Because the dual particle is a scalar, the charged defect needs to be a type of instanton in three dimensions and the monopoles are the natural objects to study.
For the case of the ABJM theory, these have been studied in various works \cite{BT,HLLLPY,KKM,Park, Ima1,Ima2,Kim:2009wb,S-JS,BKK,BP,KM}.

The purpose of this paper is to characterize the spectrum of magnetic monopole operators for various orbifolds of the ABJM model. Some of these (in the abelian orbifold case) have already been studied in various works for the special case of a single brane in toric setups  \cite{Ima1,BKK,KM,Imamura:2009hc,IK,MSparks}. This is characterized by a $U(1)^{2k}$ gauge group. The ideas presented in this work can also be applied to a large collection of possible duals to M2-branes in these setups, that have been proposed
\cite{Ueda:2008hx,Imamura:2008qs,Kim:2007ic,Hanany:2008c,Imamura:2008ji,Franco:2008um}.  A more complete analysis  has been recently completed in \cite{MSparks2}. We want to do a general analysis that includes the non-abelian orbifolds and arbitrary rank gauge groups as well. Such a program along with various of the techniques we are going to use was performed for the ABJM model
in \cite{BT,BP}. Most important for us, was the observation of \cite{TY,IK} that the moduli space for level $k$ gives collections of branes on the $\BC^4/{\BZ_{kn}\times \BZ_n}$ space, rather than the naive $\BC^4/{\BZ_k\times \BZ_n}$ quotient. In our generalization to non-abelian orbifold we will see that the pattern persists, and we get a collection of branes on $\BC^4/{\BZ_{k|\Gamma|}\times \Gamma}$, where $|\Gamma|$ is the order of the group $\Gamma$.

The main issue is to just solve for the detailed structure of the moduli space of vacua. The chiral ring will be a complete set of holomorphic coordinates on this moduli space of vacua. This is the idea of holomorphy: holomorphic operators are a complete set of order parameters to distinguish all of the different vacua of a supersymmetric theory. We are not aware of any example where this is not the case, nor of a proof of this statement in general.

 There are various parts to such an analysis. First, we will find the general solutions of the scalar field vevs that describe such a moduli space. We will show that this can be done very conveniently with the theory of representations of certain $\BC^*$ algebras associated to a quiver diagram. This is a generalization of the techniques introduced in \cite{BJL} to solve the moduli space of vacua of four dimensional theories. Such a connection with operator algebras simplifies a lot of the analysis and describes very elegantly the method of images for orbifolds of Douglas and Moore \cite{DM}. The main advantages is that one does not have to write the lagrangian of the orbifold with all of the fields explicitly, but instead one writes the lagrangian of the parent theory and imposes extra algebraic relations that make the extraction of the field content and gauge symmetries of the quotient theory manifest.

 Once we have the moduli space of vacua, there are discrete gauge identifications between the solutions that need to be addressed. To do that we need to understand how the chiral ring operators are related to the moduli space of vacua. We do this by considering the operator state correspondence and analyzing the complete set of classical BPS states of the field theory on the cylinder. These can be seen to be related to the classical moduli space of vacua in a very direct manner. The analysis of the Chern-Simons equations of motion and the quantization of gauge fluxes plays a crucial role at this stage. These classical solutions can be seen to have a natural Poisson structure on them: the complex structure of the moduli space. This lets one quantize the classical problem by holomorphic quantization. Consistency with the constraints of the Chern-Simons degrees of freedom selects the polynomial wave functions that are allowed. This provides in the end the complete set of chiral ring operator quantum numbers that are allowed. With this information we can then provide the exact topology of the moduli space of vacua of the theory, thereby generalizing various results to non-abelian orbifolds. It is clear that these techniques can be also applied in other cases.

Furthermore we can provide interesting tests of the duality of the quiver orbifold theories  with the ABJM orbifold models. Particularly, we can recover the description of how D0-brane fractionate when they arrive at a singularity of the $AdS_4\times X_6$ geometry. We see clearly how the familiar patterns of tensions expected from the local nature of the orbifold singularities of $X_6$ happen for fractional D0-branes. This ends up being intimately tied to the levels of the different Chern-Simons terms in the action. Moreover, these can be read from the quiver diagram at a glance. We will explain how this works in detail.

The paper is organized as follows. In section 2 we give an overview of the problem of computing the moduli space of vacua in three dimensions. We present a comparison with the four dimensional case to remark the importance of non-perturbative effects in 3d. In section 3 we review the orbifold construction for gauge theories and setup the problem of solving the superpotential vacuum equations as finding the irreducible representations of some quiver path algebra. In section 4 we characterize the chiral ring of these theories in terms of classical solutions to the BPS equations. In section 5 a particular example of a $\mathbb{Z}_{n}$ orbifold is solved in full detail.  We begin by explicitly showing the isomorphism between the quiver algebra and the corresponding crossed algebra. We use this to build explicitly the branes in the bulk by the method of images (this is the same as studying the general irreducible modules of the algebra). Using this prescription we describe the full  moduli space including the singular points where brane fractionation occurs.  In section 6 the previous results are generalized to non-abelian orbifolds and we show that in the non-singular locus the topology of the moduli space have the general form  $\BC^4/{\BZ_{k|\Gamma|}\times \Gamma}$. In section 7 we present a summary and a conclusion on the results obtained along with possible further directions.

\section{Moduli space problem in diverse dimensions.}\label{sec:modspace}

Let us consider a quiver gauge theory in four dimensional field theory associated to branes probing the tip of some
Calabi-Yau geometry. This is a special class of theories with gauge fields and a superpotential. The theories we are analyzing in three dimensions have this similar structure, with the extra twists of not having a Yang-Mills lagrangian, and instead having a Chern-Simons term in the action. Since the ABJM model has the same superpotential as a four dimensional model of branes at the conifold, this structure is expected to be common.

We can then compactify the four dimensional system to three dimensions and compare it to the three dimensional model with Chern-Simons terms.

The four dimensional theory reduced to three dimensions will have a Yang-Mills lagrangian for the gauge fields. In this theory the gauge coupling constant becomes large in the infrared since it has dimension $\frac{1}{2}$. The dimensional reduction of a vector multiplet from four to three dimensions contains apart from the vector potential degrees of freedom, an additional massless scalar field in the adjoint representation. This is the fourth component of the gauge field in four dimensions. We can give a vev to this component in three dimensions without breaking the supersymmetry. The off-diagonal modes become massive via a supersymmetric Higgs mechanism. Also, for the moduli space problem in four dimensions, the K\"ahler potential usually doesn't matter, so we will take it to be canonical for simplicity.
In three dimensions the K\"ahler potential is important to determine if a theory has conformal symmetry or not. All the theories we study in detail in this paper have this property anyhow, so we will not comment on this further.

These extra scalar fields coming from the vector multiplet, as long as they are massless,  can in general get vevs without breaking supersymmetry.
If we explore these vevs, we can be in a mixed Coulomb-Higgs branch, depending on the vevs of the other matter fields. This extra adjoint field, that we will call $\sigma$ increases the dimension of moduli space from $6$ real dimensions for a brane in the bulk, to seven dimensions. This is natural from the point of view of lower dimensional branes exploring some geometry. There are extra directions from the position of the brane in the dimension that is not wrapped any longer. For $N$ branes at the same locus in the four dimensional theory, a vev of this scalar field, at a generic point of the moduli space would break the gauge group to $U(1)^N$ ($\sigma$ is hermitian and can be diagonalized) . The vector field superpartners of these scalar fields will be massless. While the other off-diagonal degrees of  freedom become massive and can be integrated out. For the chiral multiplets, it is easy to show that only diagonal components remain massless also. This is because the kinetic term (for a canonical kinetic term in four dimensions) contains terms that contribute to the potential which are of the form
\begin{equation}
|[\sigma, \phi]|^2
\end{equation}
These are from the dimensional reduction of the terms with covariant derivatives in the fourth direction.
Then, in the infrared, the massless chiral fields will be decoupled from the diagonal vector fields, since they will satisfy $[A_{\mu},\phi]=0$. Therefore, the low energy effective theory has no massless charged particles under the $U(1)^{N}$ gauge group.

In this setup, we are considering the moduli space at a generic point in the Calabi-Yau geometry associated to the four dimensional theory, where  the unbroken gauge group is $U(N)$ and all moduli are in the adjoint: we expect that this low energy effective theory is like ${\cal N}=4$ SYM away from the tip of the cone. In the full theory of $N$ D3-branes on a Calabi-Yau singularity, this $U(N)$ is embedded diagonally in the quiver gauge theory, whose gauge group is $G=\prod_{i} U(N_{i})$ with $\sum_{i}N_{i}=N$, and all matter fields transform in the adjoint of this diagonal $U(N)$.

This shows that only the chiral fields $\phi$ that are mutually diagonal with the $U(N)$ are allowed. As we pointed out before, at a generic point of the moduli space, in the infrared we have a free theory for $U(1)^{N}$ vector fields and massless scalars. We want to analyze these $U(1)$ degrees of fredom carefully.
For a $U(1)$ vector field, $V_\mu$, we can dualize the field strength $F_{\mu\nu} \sim \epsilon_{\mu\nu\gamma}\partial^\gamma \theta$ to write it in terms of an electromganetic dual scalar field. The equation of motion of the free $F$ in the low energy effective field theory becomes
a Bianchi identity for this expression, and the Bianchi identity for $F$ becomes a Laplacian acting on $\theta$ that gets set to zero. This is in the procedure in the absence of sources. This dual scalars $\theta$ can also be considered to be in the adjoint of $U(1)^N$. Now, $\theta$ can also get a vev, but it is not visible in perturbation theory of the original lagrangian. A gradient of $\theta$ is visible as an electro-magnetic field, and in the original lagrangian this is in the adjoint of $U(N)$, so that one can assume that a putative non-abelian completion of $\theta$ is also in the adjoint of $U(N)$, but this is just so that we can understand that when the $\theta$ get vevs, the gauge group should be broken also.

The condition that a vev of $\theta$ is massless in the Coulomb branch can be described as $[\sigma, \theta]=0$, so that
the combined vacuum expectation values of $\theta, \sigma$ on the moduli space of vacua break the theory to $U(1)^N$ and no further. As long as we're doing the analysis in the low energy effective field theory with $U(1)^N$ symmetry, this dualization procedure can be done without much trouble. For the full non-abelian symmetry we do not know of a way to do this consistently for every case. These two extra dimensions get naturally complexified, and suggest that the moduli space of vacua for a single brane grows one extra complex direction, described by one perturbative vev $\vev{\sigma}$, and one non-perturbative vev $\vev{\theta}$ that we need to access somehow \footnote{For theories with more supersymmetry, this gives interesting topological effects \cite{SW} that let one solve for the moduli space metric exactly.}.

This can only be done non-perturbatively. Naturally $\theta$
being a scalar potential,  couples to point-like defects in three dimensions. The electric sources for $\theta$ are magnetic monopole instantons. The action for such an instanton coupling is proportional to $\theta$, but in quantum effects
it must be exponentiated: only the exponential of the action counts.

This implies that the monopole instanton can be described as a local operator inserted at the center of the monopole and it should behave as
\begin{equation}
M(x) \sim \exp( i \theta)(x)
\end{equation}
This suggests that the scalar $\theta$ is periodic. This property of monopole operators is described in detail in \cite{Polyakov}, where the dual scalar is introduced in a path integral formalism carefully.

With this information, the full moduli space of the theory is characterized by the original Calabi-Yau geometry and a $(\theta, \sigma)$ pair for each brane.

Now, let us add the Chern-Simons terms to the lagrangian. These give a topological mass to both the gauge field $V_\mu$ and $\sigma$. This means that $\theta$ also becomes a massive degree of freedom, even if we can not write an obvious lagrangian for $\theta$. This is because $\theta$ encodes the same degrees of freedom as $V_\mu$. In this situation, one expects that the vector field degrees of freedom decouple in the infrared. The vevs of monopole operators $\vev{M(x)}$ will probe the vevs of $\theta$. Also, as shown in \cite{SW}, the angle-variable $\theta$ can fiber non-trivially on the moduli space.

From now on,  we will begin in three dimensions with a Chern-Simons lagrangian, and treat the vector fields as of dimension one. The Chern-Simons coupling is marginal. If we add a SYM term to the action, we find that this is an irrelevant deformation that we can neglect in the infrared. Therefore in our following analysis of the low energy theory we can consider only the CS term. The supersymmetric Chern-Simons term adds the following coupling to the lagrangian:
\begin{equation}
\int d^3 x K D \sigma
\end{equation}
The full superfield expressions can be read in \cite{BKKS}.

So, the moduli equations describing the vacuum change. The D-term constraints are relaxed so that $\sigma$ becomes
a (background) field dependent FI term relative to the matter. Moreover, the Chern-Simmons terms give rise to a topologically massive vector field. this means that the theory does not necessarily become strongly coupled in the infrared anymore. In the infrared, such fields can be integrated out, so they should drop out of the action somehow.

 In the equation above, since $\sigma$ has no dynamical degrees of freedom left over, it
becomes a composite field. This means we do not necessarily reduce the dimension of moduli space, unless the D-term constraints have no solution for a given set of values of $\sigma$. For the ABJM theory and related models, there is a constraint between the levels of the various Chern-Simons theories that is required for this to happen \cite{MSparks}.
\begin{equation}
\sum_i \alpha_i K_i=0
\end{equation}
For $U(1)^k$ theories, $\alpha_i=1$. In general, a similar analysis shows that $\alpha_i = \hat N_i$ should be the rank of the gauge group products on a single brane moduli space,  by taking traces over the D-term constraints and summing. This constraint has a nice interpretation in terms of the diagonally embedded $U(1)$ gauge degrees of freedom: the effective Chern-Simons coupling for this diagonal field vanishes. This means that in the effective action the topological mass vanishes, and the theory requires us to include higher order terms. This is just an effective SYM action to leading order. This is the essence of the emergent SYM action from spontaneously broken conformal symmetry \cite{MP}. A topological mass for a low energy effective field vanishes, and therefore in the low energy effective action that field can not be integrated out. This keeps this direction of moduli space without it being lifted.

It also happens that in these theories, because of the Chern-Simons lagrangian, the monopole operators carry electric quantum numbers that depend on the level of the Chern-Simons pieces. This means that the non-perturbative $\theta$ vacuum expectation value should mix with the
other degrees of freedom. Since vacuum expectation values of $\theta$ also break the gauge symmetry, one can
just assume that they are fixed to some value. Under this assumption the corresponding gauge phases of the gauge group become dynamical on the other fields and can distinguish vacua. This means that on the moduli space we do not impose one D-term relation, and the corresponding phase of the associated $U(1)$ gauge group is declared to be non-gauge.
The moduli space is then for a single brane is not a standard symplectic quotient by a product group $\prod_j GL(\hat N_j,\BC)$, that would give a Calabi-Yau geometry
\begin{equation}
CY= \{\hbox{F-terms}=0 \} //\prod_j GL(\hat N_j,\BC)
\end{equation}
 but instead it is a quotient by
\begin{equation}
{\cal M}= \{\hbox{F-terms}=0 \} //\left( \prod_jGL(\hat N_j,\BC)/ GL(1,\BC)\right)
\end{equation}
So we always find a natural fibration of the moduli space over a Calabi-Yau complex manifold, and there is a natural symplectic quotient describing the CY geometry.
\begin{equation}
CY = {\cal M}// GL(1,\BC)
\end{equation}
A natural question is then what is the topology of this fibration. In particular, ${\cal M}$ has a circle action on it (the compact part of $GL(1,\BC)$), and this is the gauge phase that we allowed to stay unfixed after using our gauge freedom
on the dual scalar.  This circle is fibered over the base non-trivially. The natural periodicity of the dual scalar suggests that there might be some additional discrete freedom of these phases that is gauged. This is related to the level of the Chern-Simons theory. Thus the topology of the moduli space of vacua depends non-trivially on the Chern-Simons levels of the quiver theory (the different topologies can be understood in the toric case \cite{MSparks}). Determining this carefully is what we want to do in this paper for a variety of theories with non-abelian gauge groups, where dualizing the gauge fields is not really an option. Instead, we assume that the phases are fixed as above, and that the allowed holomorphic coordinates of the moduli space coincide with the chiral ring of the theory. We can compute the chiral ring by using other semiclassical techniques, giving us the answers we are looking for. Moreover, we see that this natural fibration makes it interesting to study the relationship between the Calabi-Yau geometry and the four dimensional complex manifold describing the moduli space of vacua of a single brane.

\section{Constructing the orbifold theories}

We want to build supersymmetric orbifold field theories of ABJM that preserve $\mathcal{N}=2$ supersymmetries in three dimensions. To do so it is best to use superspace methods to describe the lagrangian. The superspace appropriate for this level of supersymmetry is the same superspace that appears in the description of four dimensional theories with ${\cal N}=1$ supersymmetry. Therefore the usual notions of superpotential and K\"ahler potential apply for the matter action.
Because the K\"ahler potential of the ABJM model is that of a free theory, the orbifolds will have the same property.
However, the vector superfields will have a different type of action than in four dimensions: a Chern-Simons action.
The superspace actions of these vector theories have been conveniently described in \cite{BKKS}. We will use the notation
of \cite{BKKS} frequently in this paper. We will also consider in some sections the addition of a standard Super Yang Mills term to the action. This is an irrelevant deformation of the infrared field theory, but it is convenient for other purposes: one can guarantee that the gauge interactions become weak in the UV. This will preserve the supersymmetry, but will break conformal invariance.

The ABJM field theory is described most easily in $\mathcal{N}=2$ superspace in terms of a quiver diagram with some additional information that describes the interactions of the theory.  As  a quiver diagram , the ABJM theory consists of two nodes. To each node we associate a vector multiplet $V_\mu^{1,2}$ in the adjoint of $U(N_1)$, $U(N_2)$. Each of these has a Chern-Simons lagrangian with levels $k, -k$ respectively. There are four chiral matter fields $A_1, A_2$ and $B_1, B_2$. The $A$ superfields transform in the $(N_1, \bar N_2)$, and the $B$ superfields
transform in the $(\bar N_1, N_2)$ representation of the gauge group. Each of these chiral superfields have $R$-charge $1/2$ and dimension $1/2$, as it corresponds to a free scalar field in 3 dimensions.

The theory has an $SU(2)$ symmetry of rotations of the $A$ into themselves, and another
$SU(2)$ symmetry of the $B$ transforming into themselves. This manifest symmetry is an $SO(4)$ subgroup of the $SO(6)\sim SU(4)$ $R$-symmetry of the ABJM model that commutes with the manifest $SO(2)$ R-symmetry of the $\mathcal{N}=2$ superspace. In the ABJM model the scalars are in a spinor of $SO(6)$. When considered as spinors of $SO(4)$
they split into $(0,\frac 12)\oplus (\frac 12 ,0)$ representations.
The $A, B^\dagger$ can transform into each other in the ABJM theory. This mixing does not commute with the $SO(2)$ R-charge that we have singled out with our choice of ${\cal N}=2$ superspace. These extra mixings will in general be broken by our choices of the orbifold group action.

The ABJM model also has a superpotential that preserves the $SO(4)\sim SU(2)\times SU(2)$ symmetry.The field content and superpotential of the matter fields are identical to the conifold field theory \cite{KW}, except that the gauge groups are $U(N)\times U(N)$ rather than $SU(N)\times SU(N)$ and the lagrangian for the gauge degrees of freedom is different.

To preserve the $\mathcal{N}=2$ supersymmetry in an orbifold, we should choose an orbifold by a subgroup of the original $SO(6)\simeq SU(4)$ R-symmetry that   commutes with the $SO(2)$ R-charge of superspace we are preserving (it has to be embedded in the commutant). We will thus consider an orbifold by a group $\Gamma$ that sits in the $SO(4)\simeq SU(2)\times SU(2)$ that acts separately in the A and B fields. Thus, the orbifolds we are studying are classified by discrete subgroups of $SU(2)\times SU(2)$.
The problem of classification of these subgroups will not be considered here in full detail.
We will consider special subgroups that are easy to construct. These are either products of discrete subgroups of the two $SU(2)$ subgroups, or diagonal embeddings of a group into
the two $SU(2)$ subgroups. In turn,  the discrete subgroups of $SU(2)$ have an ADE classification that is well understood. Thus, we will be able to use this classification to build new quiver diagrams starting from the ABJM model using the method of images. This can be made more formal using group theory analysis and representation theory of algebras as described in \cite{BJL}. This is conveniently expressed in terms of a crossed product algebra and stating that a physical configuration is always a representation of an algebra up to isomorphism.  The equivalence up to isomorphisms is encoded in the fact that gauge theories allow gauge transformations, and physical configurations are equivalence classes under gauge transformations.
We will describe this construction in detail later in this section.

A quiver theory is usually presented as a graph with nodes and arrows. The nodes represent gauge groups, and the arrows are
interpreted as matter fields in various bifundamental representations of the gauge groups depending on the beginning and end of the arrows.

The set of arrows and nodes of a quiver can be thought of as describing some sort of matrix algebra as well (the path algebra of the quiver). Because incoming arrows and outgoing arrows are in fundamental (antifundamental) representations, we can contract them using matrix multiplication. This tends to produce composite arrows that can be thought of as composite meson fields and that also transform in bifundamental representations. The operators act on an auxiliary Hilbert space as follows. If at each node $s$ we have a gauge group $U(N_s)$, then we can build an auxiliary Hilbert space given by
\begin{equation}
{\cal H} = \oplus_s {\CV}_s
\end{equation}
where $\CV_s$ is a vector space of dimension $N_s$ in the fundamental of $U(N_s)$. All the fields of the theory can act on ${\cal H}$ and produce new elements of ${\cal H}$, because of the index structure of matrix multiplication. Under gauge transformations, ${\cal H}$ transforms in an obvious way. This can be thought of as reshuffling the basis of the $\CV_s$.
This can be done for each position in space if we want to. Here, we are indicating the matrix structure only.

We will be dealing with the scalar chiral fields and their complex conjugates and with a standard condition of reality. In mathematical terms, we are saying that we are interested in a $\BC^*$ algebra structure. In other setups it is customary to use a holomorphic path algebra only \cite{BL}, as that is the simplest way to describe the chiral ring of field theories in four dimensions. Such an algebraic approach includes the F-term equations of the field theory as part of the description of the algebra. However, the setup we need requires a slightly different take on these ideas which is why we are spending a lot of effort describing it in this slightly more elaborate way.

The
discrete symmetry of the orbifold will act on these nodes and arrows in some way, so that it preserves the action (lagrangian) of the system. This can be translated as saying that we have an automorphism of
this operator algebra that acts on ${\cal H}$, preserving some structure. For the purposes of this paper, it suffices to study symmetries that leave the nodes fixed. This is, the discrete group will not change one type of gauge field into another. More general actions can be found in various examples \cite{BL}.

To understand what the crossed product structure is, we first build the group algebra of $\Gamma$, which we will call $\BC \Gamma$. This is an algebra with a generator $e_g$ for each element $g\in \Gamma$ and any element of the algebra is a formal linear combination of
these generators with coefficients in $\BC$. The multiplication in the algebra is done as follows
\begin{equation}
e_g e_{g'}= e_{g\circ g'}
\end{equation}
which makes obvious the multiplication rule in the group. Associativity in the algebra follows from associativity of the group multiplication. Knowledge of $\BC \Gamma$ is equivalent to the knowledge of $\Gamma$. This algebra has an identity
$1=  e_1$.
A representation of $\Gamma$ of dimension $d$  is equivalent to a representation of $\BC \Gamma$ in terms of $d\times d$ matrices. This is, a map  $\mu:\BC \Gamma\to M_{d\times d}(\BC)$ that preserves all the algebraic relations (sums,  products and multiplications by scalars) and such that $\mu(1) = 1$. It is a standard result in finite group theory that all representations are unitary, and moreover they admit decompositions into direct sums of irreducibles. These are all finite-dimensional. Thus any finite dimensional representation of $\Gamma$ can be written as a sum $R= \oplus N_i R_i$, where the $R_i$ are irreducible, and the $N_i$ are the multiplicity of these irreducible representations. On each of these $R_i$, we can choose a canonical matrix representation for $\Gamma$. This is a gauge choice. This implies that the $\mu(g)$ can be assumed to be completely known and fixed by the $N_i$ labels.

If one builds an orbifold according to the prescription of Douglas and Moore \cite{DM}, we have to gauge a discrete symmetry $\Gamma$. This can be thought of as some action of $\Gamma$ on the operator algebra of a quiver diagram up to gauge transformations. This should be though of as a group action on the fields of our theory that preserve various desired structures. For example, connections should map to connections, etc.

 When acting on the gauge group (on the $V_\mu$ multiplets), we have to embed
the symmetry in the gauge group via some representation of the right dimension, characterized by a gauge transformation $\gamma(g)$ for each element $g$ of the group.

This leads to the following equation for the gauge field connection
\begin{equation}
\gamma(g) V_\mu \gamma^{-1}(g) = V_\mu
\end{equation}
This indicates the invariant nature of the gauge field under the orbifold action. Usually we ask that the discrete symmetry we are gauging does not act on the coordinates along the brane. There is another similar way to write these equations that encodes the geometric information better
\begin{equation}
D_\mu \gamma(g) (x)= 0
\end{equation}
These indicate that if we had chosen the embedding of $\gamma(g)$ to be position dependent, then the structure of the embedding is such that it is covariantly constant. We can then choose a gauge where it is constant. This produces a reduction of the holonomy group to the commutant with $\gamma(g)$. The condition on $V_\mu$ written above has exactly that interpretation: the allowed connections are those in the commutant of $\gamma(g)$.

If we interpret this as an algebraic equation for matrices, we can read this equation as if we have associated matrices $\gamma(g)$ to the elements of the group algebra $e_g$ of $\Gamma$ via a map $\mu$ as above. These can be interpreted as linear operators acting on ${\cal H}$ also. Thus, we can read the operator equations as
\begin{equation}
e_g V_\mu e_{g^{-1}} = V_\mu
\end{equation}
This is done for each gauge field that we have, with different possible $\gamma(g)$ for each.
Writing it this way we are stating that the algebraic relations are such that the algebra $\BC\Gamma$ is part of the
full algebra, rather than an external object.

The quiver algebra should also have an idempotent $\pi_s$ for each node (parametrized by $s$) in the quiver. These satisfy
\begin{equation}
\pi_s \pi_r=\delta_{rs} \pi_s
\end{equation}
This can be thought of also as the generator of the $U(1)$ 'baryon' symmetry at each node, the one that distinguishes the fundamental and the antifundamental representation of $U(N_s)$. We can recover the $\CV_s$ by projecting on the corresponding nodes with $\pi_s$, $\CV_s\simeq \pi_s {\cal H}$. These projectors are very useful objects to consider.

The fact that the $\gamma(g)$ don't permute the gauge fields into each other is expressed as follows
\begin{equation}
e_g \pi_s= \pi_s e_g \label{eq:grpaction}
\end{equation}
Moreover, because we have the direct sum decomposition in the $\mathcal{V}_s$ already spelled out, the $\pi_s$ are diagonal by blocks. Their eigenvalues are one or zero. Again, we can say that the $\pi_s$ are covariantly constant and produce a reduction of the gauge group to the gauge groups of each node.

Finally, for the matter fields, a bifundamental matter field associated to an arrow connecting nodes $s$ and $s'$ will be associated to a matrix such that
\begin{eqnarray}
\pi_r \phi^i_{ss'}&=& \delta_{rs} \phi^i _{ss'} \\
\phi^{i}_{ss'} \pi_r&=& \phi^i_{ss'} \delta_{rs'}
\end{eqnarray}
This just indicates that it is an off-diagonal matrix connecting the corresponding $\mathcal{V}_s, \mathcal{V}_{s'}$
These equations merely indicate how the fields are charged under the different gauge groups.

Also, we should impose standard hermiticity conditions as follows
\begin{eqnarray}
(\phi_{ss'}^i)^\dagger &=& \bar \phi^i_{s's}\\
\pi_s^*&=& \pi_s\\
e_g^*&=& e_{g^{-1}}
\end{eqnarray}
Notice that writing the equations in this way, we are starting to forget the labels $N_s$. This is a very convenient point of view, because what we care about are the relations in the algebra, which are independent of the values of $N_s$.
The values of $N_s$ are obtained from studying a particular representation of the algebra. Whereas if we study a general representation we can decompose it in terms of irreducibles. The nature of this decomposition is diagonalization by unitary transformations of various fields.

So far, we have an action of the original quiver algebra on ${\cal H}$, and now we have an action of $\BC \Gamma$ on ${\cal H}$ by unitary transformations. We also have the compatibility conditions $[V_\mu ,e_g]=0$, which is an algebraic representation stating that the gauge field is $\Gamma-$invariant. Notice that these equations can also be applied to the Yang-Mills curvature
\begin{equation}
[F_{\mu\nu}, e_g]=0
\end{equation}
and in general, composite fields will have definite commutation relations with the $e_g$.

One also has the invariance condition under the orbifold action for the scalar fields, given by
\begin{equation}
\gamma_s(g) \phi^i_{ss'} \gamma_{s'} (g^{-1})= R^i_j(g) \phi^j_{ss'} \label{eq:R-action}
\end{equation}
where $R$ is the action of the group $\Gamma$ on the matter fields and the embedding of $\Gamma$ in the gauge group is given by the representation $\gamma(g)=\oplus_{s}\gamma(g)_{s}$. These can also be read abstractly as
\begin{equation}
e_g \phi^i_{ss'}  e_{g^{-1}}= R^i_j (g)\phi^j_{ss'} \label{eq:phiaction}
\end{equation}
In the matrix algebra whose generators are the (vacuum expectation values of the) quantum fields of the theory, the equations of how the $e_g$ relate to each other and the fields are external constraints. Solving these equations gives a representation of the formal algebra generated by the symbols $\pi_s, e_g, \phi, V_\mu$ subject to the list of equations that
we have written above. The size of the representations are determined by the brane charges that one wants to analyze in a specific example, but these can be left undetermined without changing the nature of the algebra relations.
If one wants to look at supersymmetric vacuum solutions, then there are additional equations that indicate that we are on a vacuum manifold and these can also be interpreted in terms of representation theory of a $\BC^*$ algebra described above, with additional equations representing the $F,D$ equations of motion for the vacuum. This is not automatic. The reason why this works is that the action is of single trace type (generated by disc diagrams), so the equations of motion from the variation read as additional algebraic relations in the path algebra of the quiver.
Since the equations of motion are covariant under the action of $\Gamma$, the equations describing the conditions for vacua or the equations of motion are compatible with the action of $\Gamma$: the action of $\Gamma$ commutes with the equations of motion. The general solution will be a solution of the equations of motion of the non-orbifolded theory, and these solutions are constrained to be covariant.

One can find the most general solution of the equations describing the quiver algebra above rather directly. These can be conveniently expressed in terms of a quiver diagram also. Notice that the $\pi_s$ commute with each other. Thus they can be diagonalized simultaneously, and the $\mathcal{V}_s$ blocks provide this diagonalization.
The equations (\ref{eq:grpaction}) are also easy to solve. Since the $\pi_s$ commute with the $e_g$, then after diagonalizing the $\pi_s$ the $e$ are block diagonal in the same basis.
It follows in a straightforward way that to each node $\pi_s$ we need to associate a representation of the group $\Gamma$. We have already seen this. The representation is the embedding associated to the $\gamma$ representation in $\mathcal{V}_s$. Any such representation $\gamma_s$ of $\Gamma$ can be written
as a sum of irreducibles
\begin{equation}
\gamma_s= \oplus N_{is} R_i
\end{equation}
where the $N_{is}$ are the mutiplicities of representation $R_i$ in $\gamma_s$. These can be written in canonical form (as given by our canonical choice of matrices described previously).

To each such factor we will end up associating a residual $U(N_{is})$ gauge group. This is what we get from $V_{\mu s}$ commuting with $\Gamma$. The $V_{\mu s}$
as a matrix has to be an element of $Hom(\oplus N_{is} R_i,\oplus N_{is} R_i)$. This just states that $V_\mu$ is a matrix in the $\mathcal{V}_s$ block.  Since $V_\mu$ respects the action of  $\Gamma$, we find, following the prescription of \cite{Lawrence:1998ja}, that it is an element of
\begin{equation}
V_\mu \in Hom(\oplus N_{is} R_i,\oplus N_{is} R_i)^{\Gamma}\simeq \oplus_i Mat(N_{is}\times N_{is})
\end{equation}
this is canonically equivalent as a set to a collection of $N_{is}\times N_{is}$ matrices for each $s$. But this is the adjoint representation of $U(N_{is})$, so the $U(\sum N_{is}\dim(R_i))= U(N_s)$ connection is reduced to a subgroup that is embedded diagonally, with the $N_{is}$ providing all the important data. We associate nodes of a quiver to these gauge subgroups.
There is one node per $s$ per irreducible representation of $\Gamma$. This is, each node decomposes
into many nodes, as many nodes as there are irreducible representations of $\Gamma$.

We also need to solve the equations (\ref{eq:phiaction}). However, it becomes more obvious how to do that in the operator language. $\phi_{ss'}$ acts as a map from $\mathcal{V}_{s'}$ to $\mathcal{V}_s$. These spaces are decomposed into $\mathcal{V}_{s}=\oplus N_{is} R_i$ and $\mathcal{V}_{s'}=\oplus N_{is'} R_i$. The field $\phi$ is an operator that transforms according in a representation $R$ of the R-symmetry group. If we act on the $N_{is'} R_{i}$ subspace for fixed $i$, we obtain objects that transform in the
$R\otimes R_i$ representation of $\Gamma$, where $R$ is determined by (\ref{eq:R-action}). This is the generalization  of the Wigner -Eckart theorem in quantum mechanics to arbitrary group actions.

 What is important then is the decompositions of $R\otimes R_i$ into irreducibles. These are
 obtained from tables of products of representations
\begin{equation}
R\otimes R_i = \oplus_{r} N_r R_r\otimes R_i \simeq \oplus_{r,k} N_r N_{ri}^k R_k
\end{equation}
where the $N_r$, $N_{ri}^k$ are the multiplicities of irreducible representations of $\Gamma$ in $R$, and $R_r\otimes R_i$ respectively.

The $\phi$ are Clebsch-Gordon decompositions for these products times matrices that commute with $\Gamma$.
The arrows from $\mathcal{V}_{s'}$ to $\mathcal{V}_{s}$ split according to these rules. For each $N_{i s'} R_i$,
$N_{js} R_j$, there will be $\sum_{r}N_r N_{ri}^j$ arrows representing the possible actions of the
$\phi$ acting on $R_i$, and each of these arrows transform in the $(N_{js}, \bar N_{is'})$ representations of the group.

The quiver algebra characterized by $A, \phi, \pi, e_g$ according to the equations above is the {\em crossed product} algebra of the quiver algebra  of $A,\phi, \pi$ subject to the automorphism by an action of $\Gamma$. This setup generalizes easily to the cases where we have discrete torsion: we use the twisted algebra of $\Gamma$ by the cocycle in $H^2(\Gamma,U(1))$ that describes discrete torsion\cite{BL} (see also \cite{Douglas:1998xa,Gomis:2000ej}). This can also be generalized to cases where $\Gamma$ acts with permutations on the nodes of the original quiver. This has been worked out in some detail \cite{BL}, but a complete analysis of what happens in this situations has not been done in the general case.

The action of the new orbifold theory is the same action as that of the parent theory. However we are restricted to field configurations that are compatible with the group action of $\Gamma$ as described algebraically above.

For the ABJM model, there are various discrete subgroups of $SU(2)\times SU(2)$ that one can consider \footnote{It is important to point out at this stage that any two embeddings of $\Gamma$ in $SU(4)$ which differ by a diagonal $U(1)$ gauge transformation are considered equivalent. If we call this subgroup of gauge transformations $U(1)_{D}$, we are really embedding $\Gamma$ in $SU(4)_{R}\otimes U(1)_{D}$ and $U(1)_{D}$ acts on an element of $SU(4)_{R}$ multiplying it by a phase. This is not surprising since at the end the only thing that matters are gauge invariant quantities. The $A$ and $B$ fields are not gauge invariant.}. We will consider two cases: a $\BZ_n$ group embedded into $SU(2)\times SU(2)$, or a discrete subgroup $\Gamma\in SU(2)$. Remember that $A$ and $B$ are doublets transforming in the $(1/2,0)$ and $(0,1/2)$ representation of the global symmetry. These are the fundamental representations of both $SU(2)$.

First, we need to consider the irreducible representations $R_i$ of $\BZ_n$. These are all one dimensional and given by $R_i\sim [\eta^i]$, where $\eta= \exp(2\pi i/n)$ is a fundamental root of unity. The classification $[\eta^i]$ is the action of the generator of $\BZ_n$ on the one dimension Hilbert space.

If we let $\BZ_n$ act on a two dimensional representation of $SU(2)$, the action is characterized by a root of unity $\omega$, such that $\omega^n=1$, where
\begin{equation}
2_{SU(2)} \sim R_\omega\oplus R_{\omega^{-1}} \sim [\eta^j]\oplus [\eta^{-j}]
\end{equation}
Remember we need to act with matrices of determinant one in order to be inside $SU(2)$.

From the ABJM theory, we get that each of the two nodes, associated to $\pi_1, \pi_2$ decomposes into $n$ nodes (the irreducibles of $\BZ_n$). These can be put side by side on a graph with the same labels. The superfields $A_{1,2}$ will transform according to some value $j,-j$ (after choosing a basis where $\BZ_n$ acts diagonally), while the $B$ superfields will transform according to some value $k,-k$. We will analyze the case where $j=k$ in a quite detailed fashion later on\footnote{Note that the transformation $(A,B)\rightarrow (-A,-B)$ corresponds to a gauge transformation, hence a diagonal $\mathbb{Z}_{2n}$ action ($j=k$) is equivalent to a $\mathbb{Z}_{n}$ orbifold.}.

The new quiver will be bipartite (this is the splitting on $\pi_1,\pi_2$). The $A$ arrows will connect nodes $(1,t)$ (associated to $\pi_1, [\eta^t]$) with nodes $(2,t\pm j)$, while the $B$
will connect nodes $(2, t)$ with nodes $(1,t\pm k)$. The quiver will look as follows. We only show the $A$ and $B$ arrows of one node.

\myfig{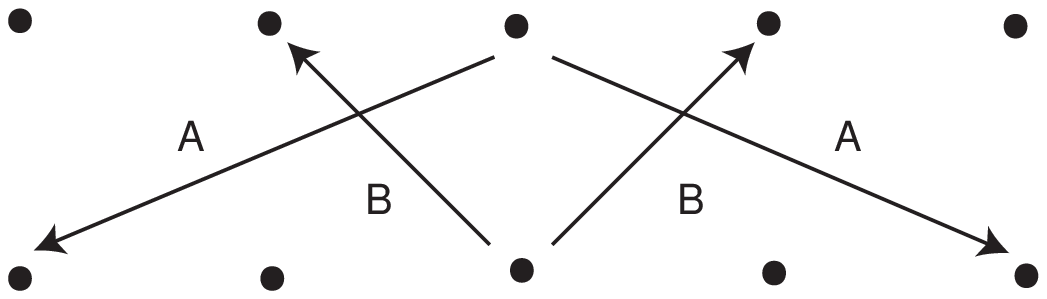}{6}{Quiver diagram for the $Q_{ABJM} /\BZ_n$ orbifold algebra. Only the fields emanating from one of the nodes inside $\pi_{1,2}$ are shown.}

Notice that the vector fields are split as follows
\begin{equation}
V^1_{\mu}\to \begin{pmatrix}
V^1_\mu[1]&&&\\
& V^1_\mu[\eta]&&\\
&&V^1_\mu [\eta^2]&\\
&&&\ddots
\end{pmatrix}
\end{equation}
where each block indicates the irreducible blocks of $R_i$ in the lagrangian. These are all one dimensional. We get a similar answer for $V^2_\mu$. With this embedding, when replacing this splitting of $V_\mu$ in the ABJM lagrangian, we find out that all the $V_\mu^1[\eta^i]$ are at level $k$, while all the
$V_\mu^2[\eta^i]$ are at level $-k$. Thus, the coupling constant is inherited in all the nodes.

We can now consider the simplest non-abelian case $\Gamma= \hat D_k$. Again, the graph is bipartite. Each of $\pi_1$ and $\pi_2$ is split into the irreducible representations of $\hat D_k$. These are the nodes of the affine $\hat D_k$ Dynkin diagram. If we tensor these with the fundamental representation (the one given by the canonical embedding in $SU(2)$), the product rules of the representations reproduces the Dynkin diagram of the affine $\hat D_k$ group.
This observation was fundamental for the understanding of dualities.

The quiver is shown in the following figure. Since the $A$ fields are chosen not to transform under $\Gamma$, they necessarily connect the same representation of the group $\Gamma$, between the nodes on the top and bottom of the figure (these are the ones associated to $\pi_1,\pi_2$).

\myfig{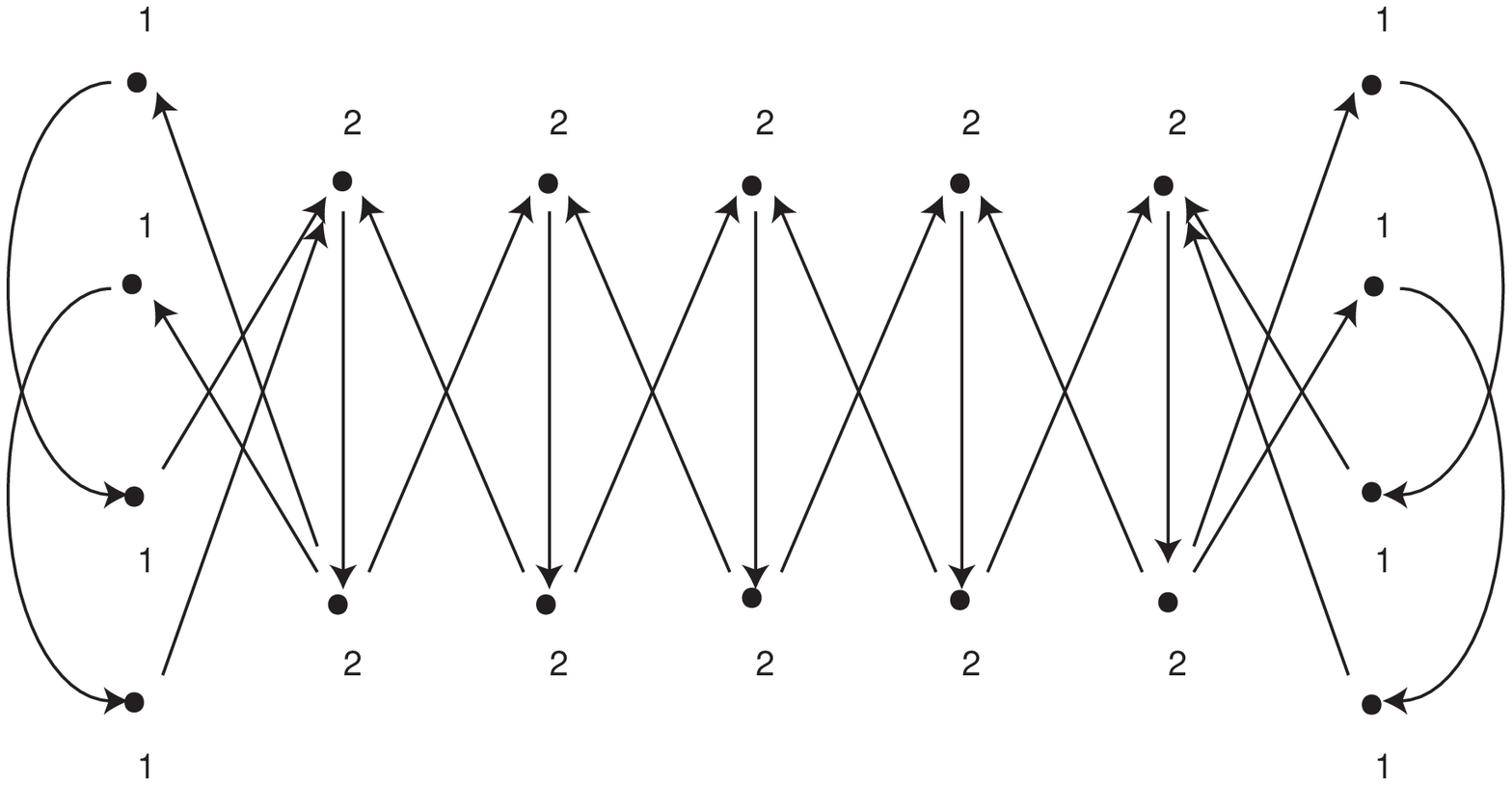}{8} {Quiver diagram for the $Q_{ABJM}/\hat D_k$ orbifold algebra. The labels on top of the representations indicate the dimension of the irreducible of $\hat D_k$ that the node is associated to. The arrows pointing downward come in pairs and transform as a doublet of the unbroken $SU(2)$ global symmetry.}

Again, we can decompose the $V_\mu^1$ according to the irreducible representations of $\hat D_k$, of the form
\begin{equation}
V^1_\mu \to \diag( V^1_\mu[R_i])
\end{equation}
The difference with the previous case is that the $R_i$ have different dimension. In the decomposition pictured above, we have that $V^1_\mu[R_i]$ is proportional to the identity of the group algebra times an $N_{1i}\times N_{1i}$ matrix.
When evaluating the action, we have to take the trace over both the $N_{1i}\times N_{1i}$ matrix and the group algebra. The level of the $R^1_i$ block is given by
\begin{equation}
k^1_{R_i} = \dim(R_i) k
\end{equation}
Similarly, we find that the levels for the $V^2_\mu$ splitting are given by
\begin{equation}
k^2_{R_i} = - \dim(R_i) k
\end{equation}
This is similar to the patterns of gauge coupling constants in orbifold theories of D-brane models with Yang-Mills interactions, where ${g_i}^{-2}= \dim (R_i) g^{-2}$.

The $\hat E$ series of discrete subgroups of $SU(2)$ is also easy to draw, it follows the same pattern of the $\hat D$ series. The example of the $\hat D$ series is enough to understand the broad patterns of behavior. The levels of the Chern-Simons orbifold theories are
$k^a_R =  \dim(R) k^a$, where $k^a$ are the levels of the parent theory.

For all of these theories the superpotential and the lagrangian are the same as those of the parent theory. The algebraic constraints imposed on the solutions (these can also be thought of as states) distinguish the theories amongst each other.

\section{Chiral operators and BPS states on the cylinder}\label{sec:chops}

Conformal field theories in $d+1$ dimensions can be characterized by the correlation functions of operators in the vacuum. In such theories there is in general an operator state correspondence that makes it possible to equate operator insertions at the origin with the spectrum of the conformal field theory compactified on a cylinder, whose base is a sphere
$S^d$. Such a compactification has a manifest $SO(d+1) \times \BR$ symmetry from isometries of the sphere and time translations.  This symmetry makes it very amenable to study the system by Hamiltonian methods. Also, the presence of a finite box implies that the spectrum of the cylinder Hamiltonian is discrete and therefore semiclassical methods can provide a good starting point to analyze the theories.

In the presence of $\mathcal{N}=2$ supersymmetries in three dimensional conformal theories, there is additionally an $SO(2)$ R-charge symmetry and a unitarity bound that makes state energies greater than or equal to their R-charge. States that saturate this inequality preserve some of the supersymmetries (they are BPS) and when quantized they generate the chiral ring of the theory. Knowledge  of the chiral ring translates directly into understanding the exact geometry of the moduli space of the theory. This point of view has been explained recently in detail in \cite{BP}. We will follow the ideas presented there to perform the calculation of the chiral ring of the orbifold theories we have considered so far. The advantage of this formulation is that it can be applied in the presence of magnetic monopole operators and that it can resolve subtle details of the geometry of moduli space.

The details follow the analysis in \cite{Bcon, BT, BP} (for other recent work, see \cite{KM}). For the ABJM theory, the complete analysis was done in \cite{BP}. Here, we can follow similar steps. The analysis is not changed substantially so long as all fields have canonical dimension and R-charge. For orbifolds this is automatic. Moreover, for orbifolds the main part of the analysis can be done in the parent theory or in the orbifold field theory without change. It is when we get to details of the solutions to the chiral ring classical states that the differences become apparent.

The first step is to go from the lagrangian formulation to the canonical quantization of the theory. This is done most simply for the matter fields in the gauge $A_0=0$. We only need to use the scalar lagrangian since we are going to look at semiclassical solution of the theory.
For the Chern-Simons fields, since the lagrangian is of first order type, the Legendre transform of the  term with first order time derivatives vanishes. We are left with a constraint whose Lagrange multiplier is $A_{0}$, hence it also vanishes. The only contributions of the gauge fields to the energy is via the terms in the lagrangian that involve the matter fields. There is also a Poisson structure for the gauge fields that is important for recovering the gauge field equations of motion from the Hamiltonian.

Since the fields are complex, the kinetic term for the matter fields is given by
\begin{equation}
K= \int_{S^2}   \tr(\Pi_\phi \Pi_{\bar \phi})
\end{equation}
where $\Pi_\phi$ is the canonical conjugate variable to the field $\phi$. If we choose a gauge $A_0\neq 0$, then one gets a different set of expressions that reflect the minimal coupling of the field to the gauge connection. The potential includes a gradient term of the fields given by
\begin{equation}
V_{\hbox{gradient}}= \int_{S^2} \tr( \mathcal{D} \phi (\mathcal{D} \phi)^{\dag})
\end{equation}
where $\mathcal{D}$ are gauge covariant derivatives along the sphere. We also have an effective mass term from the conformal coupling of the scalars to the background curvature of the sphere. In units where the sphere is of radius one, we have that this is equal to
\begin{equation}
V_{\hbox{conformal mass}} = \int_S^{2} \frac 14 \tr(\bar\phi \phi)
\end{equation}
For an s-wave mode on the sphere on a trivial gauge background, the corresponding frequency of the oscillator is given by $w^2= 1/4$, so that $w=1/2$. This reflects the fact that a free scalar field has dimension $1/2$ in three dimensions.
These expressions are independent of which orbifold we are choosing: the schematic form of the lagrangian is the same, and the dimensions of fields do not change. The interpretation of the group algebra constraints change between theories,
but at this level they eliminate fields and their canonical conjugates in pairs.

There are additional terms in the lagrangian from the potential of the theory. If we use the superspace appropriate for ${\cal N}=2$ supersymmetry in three dimensions (the same standard superspace of four dimensions), then it is convenient to write the interaction potential as
\begin{equation}
V_{\hbox{potential}} \sim \tr([\sigma,\phi][\sigma,\bar\phi]) + |W_\phi|^2
\end{equation}
which makes manifest the fact that it is a sum of squares. Again, this expression is independent of the orbifold constrains. The components of $\phi$ that can be non-zero vary between models, but the action and the Hamiltonian is identical to the one of the ABJM model. These are simply constraints on the fields.

Each of the chiral scalar fields has $R$-charge one half, as inherited from the parent ABJM theory. This means that the $R$-charge is given by
\begin{equation}
Q_R = \sum_\phi\int_{S^2}  \tr(\frac{i}{2} \Pi_\phi \phi -\frac {i}2\Pi_{\bar\phi}\bar\phi)
\end{equation}
This generates R-charge rotations by Poisson brackets
\begin{equation}
\delta_R \phi \sim \{Q_R,\phi\}_{PB}
\end{equation}
If we consider the BPS unitary inequality $H-Q\geq 0$, we can look for solutions that saturate this inequality. It is easy to show that
\begin{equation}
K+V_{\hbox{conformal mass}}-Q_R= \int  \tr(\Pi_\phi\Pi_{\bar\phi} +\frac 14 \bar\phi\phi)- Q_R =
\int \tr(\left|\Pi_{\bar\phi} - \frac i2 \phi\right|^2)
\end{equation}
So that when we consider $H-Q_R=0$, we find that $H-Q_R$ is a sum of squares. Each of these has to vanish.

These result in the following sets of equations
\begin{eqnarray}
\mathcal{D} \phi=0
\\
 W_\phi= [\sigma,\phi]=0\\
\Pi_{\bar\phi} =\dot \phi = \frac i 2 \phi
\end{eqnarray}
The first equation says that the scalar field is covariantly constant on the sphere. These equations imply that $\phi$  is spherically symmetric. If we supplement these conditions with the equations of motion of the gauge field ,the equation of motion of $A_{0}$ implies that the gauge field curvature $F_{\theta\varphi}$ is also covariantly constant in the sphere. The $A_{\theta}$ and $A_{\varphi}$ equations imply $F^{i0}=0$, which in our gauge choice, reduces to $\dot{A}_{i}=0$.
The second line above indicates that the interaction potential vanishes.

This is the condition that needs to be satisfied by a solution of the moduli space of vacua of the theory on flat space in order to have a supersymmetric vacuum. Since the field is covariantly constant on the sphere, this implies that the field is constant in an appropriate gauge as an initial condition.  The first order equations indicate that the field remains constant after evolution in the gauge $A_0=0$, so the initial gauge condition is compatible with the gauge $A_0=0$ that we chose previously.

Putting these results together, we find that the BPS classical configuration are classically in  correspondence with points in the moduli space of vacua of the theory. Notice that we have to be careful because we have not completely analyzed the gauge redundancies and how they affect this correspondence. This is especially important when  quantizing the results. At the classical level the gauge redundancy of solutions is not as important to describe the dynamics.

If we include the Gauss' law constraints, (the equation of motion of $A_0$), we find that the magnetic field is covariantly constant and given by the current of schematic form
\begin{equation}
k F =  (\phi \Pi_\phi -   \bar \Pi_\phi \bar \phi) \sim  \frac i2  \phi\bar\phi
\end{equation}
where we have assumed $\phi$ is in the fundamental and we have used the BPS equations of motion. For the antifundamental, signs and ordering are reversed. Both contribute. Remember that $F$ is a matrix, as well as $\Pi, \phi$. To take into account both possibilities,
this can be written as the following matrix equation
\begin{equation}
kF_v= \frac i2 \pi_v  [\phi,\bar\phi]
\end{equation}
One of the products will be zero in the quiver algebra because of the projector $\pi_v$.
The notation includes implicit matrix multiplication, which also affects the ordering of the fields.
These covariantly constant solutions of the magnetic field are also solutions of the Yang-Mills equation in two dimensions. The magnetic fluxes are quantized at the classical level as originally shown by Atiyah and Bott \cite{AB}.

So the program is clear: we need to first evaluate the classical vacuum equations of the field theory in flat space. We then need to impose these as initial conditions of  the theory. The manifold of initial conditions has a Poisson structure. It is induced from the first order equations of motion treated as constraints.  Since $\Pi_\phi \sim \bar \phi$, we see that $\bar \phi$ becomes canonically conjugate to $\phi$ in a well defined sense.

This means that the Poisson structure of initial conditions makes the holomorphic variables a complete set of commuting coordinates. This lets us perform a holomorphic quantization of the moduli space of vacua: wave functions are holomorphic wave functions. These are supplemented by a measure that we will not determine \footnote{These measures can be calculated in a semiclassical limit \cite{Bem, Bcor, Bcot,BT}, and based on the structures found generalizations can be made to other setups \cite{BHart}. The calculated measure can be used to match other calculations that can be done at weak coupling in three dimensional theories\cite{BT}, but it is not understood how to calculate these measures at strong coupling.}

These wave functions also need to be single valued, which places constraints on them. These holomorphic wavefunctions end up describing the full structure of the chiral ring in the ABJM case \cite{BT,BP}. This lets one study the exact topology of the moduli space of vacua: the chiral ring is assumed to be the complete set of order parameters classifying the vacua of a supersymmetric theory.

There are two ways to proceed now. We can either analyze the quiver theory of the orbifold or we can analyze the theory in the parent theory and impose the projection conditions,  and recover the same information. Both ways of proceeding will give the same answer in this case. We will show how this works in a particular example in a lot of detail by working directly in the orbifold theory. We will then see what implications the second formulation has in the case of non-abelian orbifolds where it is more convenient.

\section{A quiver example in complete detail}

\subsection{The BKKS example}
%%%%%%%%%%%%%%%%%%%%%%%}
We consider a modification of the ABJM theory \cite{ABJM} with $G=\prod_{i=1}^{2n}U(N_{i})$, described by Benna et al. in \cite{BKKS}. We will call this model the BKKS model for simplicity. The field content and conventions
are mostly from \cite{BKKS}. The quiver is given by
\myfig{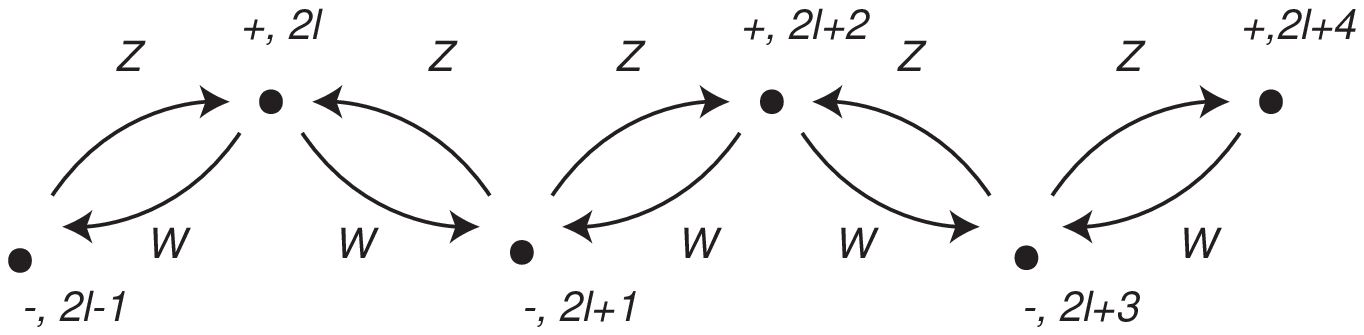}{8}{Quiver diagram the BKKS orbifold. The nodes are numbered and are given a sign: the sign of the Chern-Simons level.  }

The orbifold acts on the $A^{1,2}$ and $B^{1,2}$ superfields by a $\BZ_n$ action. The generator of $\BZ_n$ $g$ acts by sending
\begin{eqnarray}
A^{1} &\to& \eta^{1/2} A^1\\
B^{1}&\to &\eta^{1/2} B^1\\
A^2&\to & \eta^{-1/2} A^2\\
B^{2}&\to &\eta^{-1/2} B^2
\end{eqnarray}
where $\eta= \exp( 2\pi i/n)$. Notice that this is a $\BZ_n$ action, because $g^n$ acts
as the gauge transformation $(-1)$. Indeed, the gauge transformations let us do any of the following identifications
\begin{eqnarray}
A^{1} &\to& \exp(i\phi)\eta^{1/2} A^1\\
B^{1}&\to &\exp(-i\phi) \eta^{1/2} B^1\\
A^2&\to & \exp(i\phi) \eta^{-1/2} A^2\\
B^{2}&\to &\exp(-i\phi)\eta^{-1/2} B^2
\end{eqnarray}
with the same field content.
There is one simple choice of phases as follows
\begin{eqnarray}
A^1 &\to& \eta A^1\nonumber \\
B^1 &\to & B^1\nonumber\\
A^2 &\to & A^2\nonumber \\
B^2 &\to& \eta^{-1} B^2 \label{eq:orb2}
\end{eqnarray}
which shows more clearly that we have a $\BZ_n$ action, but it is less symmetric. These choices are equivalent.

The graph is Identified at both ends: node 0 and node $2n$ are the same. From the picture the superfields transform as
\begin{eqnarray}
\mathcal{Z}_{2l}\rightarrow U_{2l+1}\mathcal{Z}_{2l}U_{2l}^{\dagger},\nonumber\\
\mathcal{W}_{2l}\rightarrow U_{2l}\mathcal{W}_{2l}U_{2l+1}^{\dagger},\nonumber\\
\mathcal{Z}_{2l-1}\rightarrow U_{2l-1}\mathcal{Z}_{2l-1}U_{2l}^{\dagger},\nonumber\\
\mathcal{W}_{2l-1}\rightarrow U_{2l}\mathcal{W}_{2l-1}U_{2l-1}^{\dagger}.
\end{eqnarray}
And their components in the superspace expansion are
\begin{eqnarray}
\mathcal{Z}_{l}=Z_{l}+\sqrt{2}\theta \zeta_{l}+\theta^{2}F_{l},\nonumber\\
\mathcal{W}_{l}=W_{l}+\sqrt{2}\theta \omega_{l}+\theta^{2}G_{l},\nonumber\\
\bar{\mathcal{Z}}_{l}=Z^{\dagger}_{l}-\sqrt{2}\bar{\theta} \zeta^{\dagger}_{l}-\bar{\theta}^{2}F^{\dagger}_{l},\nonumber\\
\bar{\mathcal{W}}_{l}=W^{\dagger}_{l}-\sqrt{2}\bar{\theta} \omega^{\dagger}_{l}-\bar{\theta}^{2}G^{\dagger}_{l}.
\end{eqnarray}
As we are interested in the moduli space then, the relevant terms in the action are those of the scalars (we omit terms that involve fermions)
\begin{eqnarray}
\mathcal{S}_{CS}&=&\frac{\kappa}{4\pi}\int d^{3}xTr\Big[\sum_{l=1}^{n}(\mathcal{L}_{CS}(V_{2l-1})-\mathcal{L}_{CS}(V_{2l})-4D_{2l-1}\sigma_{2l-1}+4D_{2l}\sigma_{2l})\Big],\nonumber\\
\mathcal{S}_{mat}&=& \int d^{3}xTr\Big[\sum_{l=1}^{n}
|\nabla Z_l|^2+|\nabla W_{l}|^2
\nonumber \\ && (F_{2l-1}^{\dag}F_{2l-1}+Z_{2l-1}^{\dag}D_{2l-1}Z_{2l-1}-Z_{2l-1}^{\dag}Z_{2l-1}D_{2l}\nonumber\\
&-&Z_{2l-1}^{\dag}\sigma_{2l-1}^{2}Z_{2l-1}-Z_{2l-1}^{\dag}Z_{2l-1}\sigma_{2l}^{2}+2Z_{2l-1}^{\dag}\sigma_{2l-1}Z_{2l-1}\sigma_{2l}\nonumber\\
&+&G_{2l-1}^{\dag}G_{2l-1}+W_{2l-1}^{\dag}D_{2l}W_{2l-1}-W_{2l-1}^{\dag}W_{2l-1}D_{2l-1}\nonumber\\
&-&W_{2l-1}^{\dag}\sigma_{2l}^{2}W_{2l-1}-W_{2l-1}^{\dag}W_{2l-1}\sigma_{2l-1}^{2}+2W_{2l-1}^{\dag}\sigma_{2l}W_{2l-1}\sigma_{2l-1}\nonumber\\
&+&F_{2l}^{\dag}F_{2l}+Z_{2l}^{\dag}D_{2l+1}Z_{2l}-Z_{2l}^{\dag}Z_{2l}D_{2l}\nonumber\\
&-&Z_{2l}^{\dag}\sigma_{2l+1}^{2}Z_{2l}-Z_{2l}^{\dag}Z_{2l}\sigma_{2l}^{2}+2Z_{2l}^{\dag}\sigma_{2l+1}Z_{2l}\sigma_{2l+1}\nonumber\\
&+&G_{2l}^{\dag}G_{2l}+W_{2l}^{\dag}D_{2l}W_{2l}-W_{2l}^{\dag}W_{2l}D_{2l+1}\nonumber\\
&-&W_{2l}^{\dag}\sigma_{2l}^{2}W_{2l}-W_{2l}^{\dag}W_{2l}\sigma_{2l+1}^{2}+2W_{2l}^{\dag}\sigma_{2l}W_{2l}\sigma_{2l+1})\Big]\nonumber\\
\mathcal{S}_{pot}&=& \frac{L}{2}\int d^{3}xTr\Big[\sum_{l=1}^{n}(F_{2l-1}W_{2l}Z_{2l}W_{2l-1}+Z_{2l-1}W_{2l}F_{2l}W_{2l-1}\\&+& Z_{2l-1}G_{2l}Z_{2l}W_{2l-1}+Z_{2l-1}W_{2l}Z_{2l}G_{2l-1})\nonumber\\
&-&\sum_{l=1}^{n}(F_{2l}W_{2l}Z_{2l+1}W_{2l+1}+Z_{2l}W_{2l}F_{2l+1}W_{2l+1}+Z_{2l}G_{2l}Z_{2l+1}W_{2l+1}+Z_{2l}W_{2l}Z_{2l+1}G_{2l+1})\nonumber\\
&+&\sum_{l=1}^{n}(F_{2l}^{\dag}W_{2l+1}^{\dag}Z_{2l+1}^{\dag}W_{2l}^{\dag}+Z_{2l}^{\dag}W_{2l+1}^{\dag}F_{2l+1}^{\dag}W_{2l}^{\dag}+Z_{2l}^{\dag}G_{2l+1}^{\dag}Z_{2l+1}^{\dag}W_{2l}^{\dag}+Z_{2l}^{\dag}W_{2l+1}^{\dag}Z_{2l+1}^{\dag}G_{2l}^{\dag})\nonumber\\
&-&\sum_{l=1}^{n}(F_{2l-1}^{\dag}W_{2l-1}^{\dag}Z_{2l}^{\dag}W_{2l}^{\dag}+Z_{2l-1}^{\dag}W_{2l-1}^{\dag}F_{2l}^{\dag}W_{2l}^{\dag}+Z_{2l-1}^{\dag}G_{2l-1}^{\dag}Z_{2l}^{\dag}W_{2l}^{\dag}+Z_{2l-1}^{\dag}W_{2l-1}^{\dag}Z_{2l}^{\dag}G_{2l}^{\dag})\Big]\nonumber
\end{eqnarray}
Where $2n+1\sim 1$ and $0\sim 2n$. Solving the equations for the auxiliar fields gives

\begin{eqnarray}
F_{2l-1}^{\dag}=\frac{L}{2}\Big(W_{2l-1}Z_{2l-2}W_{2l-2}-W_{2l}Z_{2l}W_{2l-1}\Big)\nonumber\\
F_{2l}^{\dag}=\frac{L}{2}\Big(W_{2l}Z_{2l+1}W_{2l+1}-W_{2l-1}Z_{2l-1}W_{2l}\Big)\nonumber\\
G_{2l-1}^{\dag}=\frac{L}{2}\Big(Z_{2l-2}W_{2l-2}Z_{2l-1}-Z_{2l-1}W_{2l}Z_{2l}\Big)\nonumber\\
G_{2l}^{\dag}=\frac{L}{2}\Big(Z_{2l+1}W_{2l+1}Z_{2l}-Z_{2l}W_{2l-1}Z_{2l-1}\Big)\nonumber\\
\sigma_{2l}=\frac{1}{4K}\Big[Z_{2l-1}^{\dag}Z_{2l-1}-W_{2l-1}W_{2l-1}^{\dag}+Z_{2l}^{\dag}Z_{2l}-W_{2l}W_{2l}^{\dag}\Big]\nonumber\\
\sigma_{2l-1}=\frac{1}{4K}\Big[Z_{2l-1}Z_{2l-1}^{\dag}-W_{2l-1}^{\dag}W_{2l-1}+Z_{2l-2}Z_{2l-2}^{\dag}-W_{2l-2}^{\dag}W_{2l-2}\Big]
\end{eqnarray}

\subsection{The moduli space}

As was described in the previous section, we need to compute the moduli space of the theory.
After this is calculated, we need to impose the equations of motion of the gauge fields and the flux quantization. First we need to solve the equations $F=F^\dagger= G=G^\dagger=0$  and
\begin{eqnarray}
\sigma_{2l-1}Z_{2l-1}-Z_{2l-1}\sigma_{2l}=0,\nonumber\\
\sigma_{2l}W_{2l-1}-W_{2l-1}\sigma_{2l-1}=0,\nonumber\\
\sigma_{2l+1}Z_{2l}-Z_{2l}\sigma_{2l}=0,\nonumber\\
W_{2l}\sigma_{2l+1}-\sigma_{2l}W_{2l}=0.
\end{eqnarray}
These equations describe the minimum of the potential. We will show how this is a lot easier by constructing the algebra of the quiver, rather than trying to solve them as they appear above.

So, we define the set of projectors $\{\pi_{i}\}_{i=1,\ldots,2n}$ which are associated with the nodes of the quiver and which satisfy $\pi_{i}\pi_{j}=\delta_{ij}\pi_{j}$. Since the quiver has a $\BZ_n$ symmetry of cyclic permutation of $+$ nodes and $-$ nodes, it is convenient to define the monomials
\begin{eqnarray}
\varsigma^+&=\sum_{i=1}^{n}\eta^{  i}\pi_{2i}\qquad& \eta=e^{\frac{2 i\pi}{n}}\\
\varsigma^-&=\sum_{i=1}^{n}\eta^{(2 i+1)/2 }\pi_{2i+1}&\end{eqnarray}
The $\varsigma^+$ and $\varsigma^-$ transform by phases when we act with the permutation of the nodes.

Clearly we can invert this formulae
\begin{eqnarray}
\pi_{2k}&=&\frac{1}{n}\sum_{j=1}^{n}\eta^{- jk}(\varsigma^+)^{j}\\
\pi_{2k+1}&=&\frac{1}{n}\sum_{j=1}^{n}\eta^{-(2j+1) k/2 }(\varsigma^-)^{j}
\end{eqnarray}
Notice also that $(\varsigma^+)^n = \pi^+$, and $(\varsigma^-)^n= - \pi^-$, these are the projectors on the even/odd nodes respectively, and that $\varsigma^+\varsigma^-=\varsigma^-\varsigma^+=0$.

Consider now an algebra with two projectors $\pi^+, \pi^-$ and a group $\BZ_n$ generated by $g$ (so that $e_g^n=1$), with the relations $[e_g,\pi]=0$. It is easy to see that this algebra is equivalent to the one generated by $\varsigma^+, \varsigma^-$, with the following identifications
\begin{equation}
e_{g}= \varsigma^++\eta^{1/2}\varsigma^-\qquad,(\varsigma^+)^n=\pi^+, (\varsigma^-)^n=- \pi^-
\end{equation}

Then, we put the $Z$ and $W$ operators in some element of the path algebra, say $\xi$, such that
\begin{equation}
\pi_{2l}\xi=0\qquad \pi_{2l+1}\xi=\xi \pi_{2l}+\xi \pi_{2l+2}\qquad \pi_{2l+1}\xi \pi_{2l}=Z_{2l}\qquad \pi_{2l+1}\xi \pi_{2l+2}=Z_{2l+1}
\end{equation}
Likewise
\begin{equation}
\pi_{2l+1}\omega=0\qquad \pi_{2l}\omega=\omega \pi_{2l+1}+\omega \pi_{2l-1}\qquad \pi_{2l}\omega \pi_{2l+1}=W_{2l}\qquad \pi_{2l}\omega \pi_{2l-1}=W_{2l-1}
\end{equation}
We should split these into their even and odd part, as follows
\begin{eqnarray}
\sum_{l=0}^{n-1}\pi_{2l+1}\xi P\pi_{2l}=\xi_{e}&\qquad &\sum_{l=0}^{n-1}\pi_{2l+1}\xi \pi_{2l+2}=\xi_{o}\\ \sum_{l=0}^{n-1}\pi_{2l}\omega \pi_{2l+1}=\omega_{e}&\qquad& \sum_{l=0}^{n-1}\pi_{2l}\omega \pi_{2l-1}=\omega_{o}
\end{eqnarray}
After using these symbols and the formal algebra manipulations, we can express all F-term conditions as
\begin{eqnarray}\label{fterm}
\omega_{e}\xi_{e}\omega_{o}=\omega_{o}\xi_{e}\omega_{e}\nonumber\\
\omega_{e}\xi_{o}\omega_{o}=\omega_{o}\xi_{o}\omega_{e}\nonumber\\
\xi_{o}\omega_{o}\xi_{e}=\xi_{e}\omega_{o}\xi_{o}\nonumber\\
\xi_{o}\omega_{e}\xi_{e}=\xi_{e}\omega_{e}\xi_{o}
\end{eqnarray}
Notice that in this formulation, we have set up the following matrices made of the $Z$
\begin{equation}
\xi_e = \begin{pmatrix}
0&0 & 0&0& \cdots&0 &0 &Z_{2n}\\
0&0&0&0&\cdots&0&0&0\\
0&Z_{2}&0&0&\cdots&0&0&0\\
0&0&0&0&\cdots&0&0&0\\
0&0&0&Z_{4}&\cdots&0&0&0\\
\vdots&\vdots&\vdots&0&\ddots&0&\cdots&0\\
0&0&0&0&\cdots&Z_{2n-2}&0&0\\
0&0&0&0&\cdots&0&0&0
\end{pmatrix} \qquad \xi_o= \begin{pmatrix}
0&Z_{1} & 0&0& \cdots &0\\
0&0&0&0&\cdots&0\\
0&0&0&Z_{3}&\cdots&0\\
\vdots&\vdots&\vdots&0&\ddots&\vdots\\
0&0&0&0&\cdots&Z_{2n-1}\\
0&0&0&0&\cdots&0
\end{pmatrix}
\end{equation}
so that they are off-diagonal connecting the various vector spaces that we have set up in each node. The multiplications with projectors encode just which off-diagonal blocks are occupied and which are empty. We have done something similar with the $W$, and called it $\omega$.

On the other hand, the D-term conditions are expressed as
\begin{eqnarray}\label{dterm}
[\Sigma,\omega_{e}]=[\Sigma,\xi_{e}]=[\Sigma,\omega_{o}]=[\Sigma,\xi_{o}]=0
\end{eqnarray}
where
\begin{eqnarray}\label{sigma}
\Sigma=\frac{1}{4K}\Big( \xi_{o}\xi_{o}^{\dag}+\xi_{e}\xi_{e}^{\dag}-\omega_{o}^{\dag}\omega_{o}-\omega_{e}^{\dag}\omega_{e}+\xi_{o}^{\dag}\xi_{o}+\xi_{e}^{\dag}\xi_{e}-\omega_{o}\omega_{o}^{\dag}-\omega_{e}\omega_{e}^{\dag}\Big)
\end{eqnarray}

It is easy to check that with $e_{g}= \varsigma^++\eta^{1/2}\varsigma^-$, $e_{g^{-1}}=e_g^{n-1}$,
$A^1=\xi_e$, $A^2=\xi_o$, $B^1= \omega_o$, $B^2=\omega_e$, then the quiver algebra spanned by the variables $Z_{l}$ and $W_{l}$ is identical to the crossed product algebra of the $\BZ_n$ orbifold. That is $\mathcal{A}\boxtimes \mathbb{Z}_{n}$, where $\mathcal{A}$ is the ABJM $\BC^*$ algebra spanned by $A,B$. The product between elements $a\rtimes e_{g},a'\rtimes e_{g'}\in\mathcal{A}_{c}\boxtimes \mathbb{Z}_{n}$ is given by
\begin{eqnarray}
(a\rtimes e_{g})(a'\rtimes e_{g'})=ae_{g}a'e_{g^{-1}}\rtimes e_{g}e_{g'}.
\end{eqnarray}
Moreover, the equations of motion describing the moduli space are the same equations of motion that one would obtain for the ABJM model. So not only do we recover the off-shell crossed product algebra from the quiver, we can also recover the equations of motion that describe the moduli space in the parent theory when we impose that we are in a vacuum configuration.

These equations are also algebraic in nature: they are matrix equations which involve only sums and matrix multiplications, so the theory of algebra representations can help solve the problem.
In particular, notice that the ABJM algebra $\mathcal{A}$ is a subalgebra of the crossed product algebra. Thus any representation of the crossed product algebra (with vacuum constraints) is automatically a representation of the ABJM model (with vacuum constraints). Since the vacuum constraints are equivariant, we can build repesentations of the crossed product algebra by inducing representations of the orbifold vacuum solutions from solutions (representations) of the ABJM model. This essentially reduces to the method of images in orbifold setups.

\subsubsection{The regular representation}

The idea now is to build the general representation of the vacuum equations of the
ABJM orbifold models from solutions of the ABJM theory for a generic case.
The structure that we need resembles the analysis of four dimensional theories very closely.
This has been discussed in \cite{MSparks}, where it is observed that the dimension of the moduli space for a single brane is one complex dimension bigger than in the case of four dimensional theories, and that the extra dimension is fibered over a base
which is the moduli space of the associated four dimensional quantum field theory. Thus, we
need to explore how this structure can be analyzed in detail and how it plays a role in our understanding of the system.

Let us begin with the ABJM vacuum representations. It has been shown in \cite{MSparks, BT,BP} that the $U(N)\times U(N)$ ABJM model vacuum solutions have a decomposition into $N$ copies of the $U(1)\times U(1)$ model: this is a direct sum of two dimensional representations of the algebra $\mathcal{A}$. For the $U(1)\times U(1)$ model, we can choose $A^{1,2}, B^{1,2}$ to be parametrized by arbitrary complex numbers.

Thus the general solution will be of the form of block-diagonal matrices as follows
\begin{equation}
A^{1,2} \sim \diag( a^{1,2}_i)\otimes \begin{pmatrix} 0&1\\
0&0\end{pmatrix}\qquad B^{1,2}\sim \diag( b^{1,2}_i)\otimes \begin{pmatrix} 0&0\\
1&0\end{pmatrix}
\end{equation}
There is a $(U(1)\times U(1))^N$ block of continuous gauge transformations that preserve this block decomposition acts as follows
\begin{eqnarray}\label{gaugediag}
a^{1,2}_i &\to& \exp( i\phi_i) a_i^{1,2}\\
b^{1,2}_i &\to& \exp(-i\phi_i) b^{1,2}_i
\end{eqnarray}
where $\phi_i = \chi^1_i-\chi_i^2$ is a sum of two phases, one in each one of the two $U(1)$ gauge groups associated to an eigenvalue. This is the 'unbroken' gauge group associated to each of the image branes. This is not the same as the unbroken gauge group of the configuration, although it seems similar. This is the gauge freedom of defining the basis as `eigenvalues' of the $A,B$ matrices. This gets frozen when we act with the group $\Gamma$ and require that
$\Gamma$ commute with the $A$ in particular ways, so that the $a_{i}$'s are related to each other. Some of these phases can survive as the unbroken gauge group of the orbifold theory.

We have to be careful with these gauge transformations. As we have noticed before, the
group of automorphisms of the orbifold might close onto a gauge transformation. This will be very important for us later on when we discuss the structure of singularities. Because of this, we need objects that have less gauge freedom to tie the analysis down.

Consider for example the composite mesons $A^s B^t+B^tA^s$ where $s,t\in\{1,2\}$. The sum is there because we are using matrix multiplication and we think of these as matrices (operators) on a Hilbert space ${\cal H}$, so the order of multiplication matters. The coordinates of these on a $U(1)\times U(1)$ brane are
\begin{equation}
{\mW}^{s,t}= A^s B^t+B^tA^s= \begin{pmatrix} a^sb^t&0\\
0&a^sb^t\end{pmatrix}
\end{equation}
where we are keeping the convention of having one vector space of dimension one for each node in the quiver (this is what the $U(1)\times U(1)$ indicates us to do).
The operators ${\mW}^{s,t}$ generate the center of the conifold algebra $\mathcal{A}_{c}$
\begin{equation}
\mathcal{ZA}_{c}=\langle {\mW}^{s,t}\rangle
\end{equation}
hence these matrices are diagonal and proportional to the identity in irreducible representations. The proportionality constant is complex.

Moreover, these are
gauge invariant under the $U(1)\times U(1)$ group.
Variables like this generates the center of the ABJM $\BC^*$ algebra. Other examples
are
\begin{equation}
A^1 (A^1)^\dagger+ (A^1)^\dagger A^1\simeq  \begin{pmatrix} |a^1|^2&0\\
0&|a^1|^2\end{pmatrix}
\end{equation}
which is clearly hermitian.

The gauge transformation (\ref{gaugediag}) does not affect these diagonal variables. Thus on these objects the action of the $\BZ_n$ algebra can be defined unambiguously (not up to a gauge transformation).

For example, as mentioned earlier, the variables $\mW^{s,t}$ are holomorphic, and the relations between them are those of the conifold
\begin{equation}
\mW^{1,1} \mW^{2,2}= \mW^{1,2} \mW^{2,1}
\end{equation}
Indeed, if we were solving the problem in four dimensions, this would completely characterize the representation theory content of points in the moduli space. This is what one does for the conifold field theory of Klebanov and Witten \cite{KW} and was analyzed using these techniques in \cite{Bcon}. The idea of using the center of the algebra to describe the moduli spaces of branes was developed in \cite{BJL}, but it was used in the holomorphic context only. In this case we need to also consider the real structure that is imposed on us from some of the equations describing the moduli space of vacua.

Also, the traces of $\mW^{[r]}$ are gauge invariant polynomials in the field theory. These would describe the (mesonic) chiral ring operators of the conifold theory in four dimensions and their vevs parametrize the moduli space of vacua.
These same polynomials form part of the chiral ring of the three dimensional field theory as well. However, there are other non-perturbative contributions that complete the chiral ring and are magnetic monopole operators. Without them one cannot understand the full moduli space. One would get the same results as the four dimensional theory.
Our purpose is to address these non-perturbative operators systematically later on.

Given these holomorphic $\mW$ variables, it is natural to consider how the $\BZ_n$ orbifold acts on them. We clearly see that
\begin{eqnarray}
\mW^{1,1} & \to& \eta \ \mW^{1,1}\\
\mW^{1,2} &\to & \mW^{1,2}\\
\mW^{2,1} &\to &\mW^{2,1}\\
\mW^{2,2}&\to &\eta^{-1} \mW^{2,2} \label{eq:orbactionz}
\end{eqnarray}
These types of orbifolds of the conifold in four dimensions have been analyzed in the work  \cite{OT}. Here we give a more complete algebraic characterization of various features of the Calabi-Yau geometry, from the point of view of algebra representations.

The center of the orbifold algebra $\mathcal{A}\boxtimes \mathbb{Z}_{n}$ is generated by the elements of $\mathcal{ZA}$ which are invariant under the action of $\mathbb{Z}_{n}$. This happens often.
Given that the rephasings by $\eta$ are to become gauged, the new set of invariants is given by $\mW^{1,2}, \mW^{2,1}$ and
\begin{eqnarray}
\mU&= &(\mW^{1,1})^n\\
\mV&=&(\mW^{2,2})^n\\
\mZ&= & \mW^{1,1} \mW^{2,2} =\mW^{1,2} \mW^{2,1}
\end{eqnarray}
notice that the variable $\mZ$ becomes redundant because of the original conifold equations.
The new relations between the variables is
\begin{equation}
\mU \mV= (\mW^{1,2} \mW^{2,1})^n
\end{equation}
However, we can understand how these variables describe the moduli space a lot better if we think of them in terms of the representation theory of the orbifold algebra.

What we need to do now is understand how the representations of the moduli space algebra can be characterized by these numbers. In particular, we can always choose a gauge where the $\mW$ are diagonal. Now we want to analyze how to put various representations together into representations of the crossed product algebra.

Making $\mW$ diagonal reflects a choice of basis on our representation space ${\cal H}$.
In this basis, if the $\mW$ are generic, they are invertible and any element of the group action changes at least one of the $\mW$ variables. This  is, it is associated to an orbit where the subgroup that leaves the point fixed is the trivial one.
If our basis is labeled by the eigenvalues of $\mW$, we have that
\begin{equation}
\mW^{i,j} \ket{w} = w^{i,j} \ket{w}
\end{equation}
where the $w^{i,j}$ are now the eigenvalues. Given one such $\ket w$, we can act with the group element $g$ to find that
\begin{equation}
\mW^{i.j} (e_g \ket{w}) = e_g (\mW^{i,j})^g \ket{w}= e_g (w^{ij})^g \ket{w}= (w^{ij})^g (e_g \ket w)
\end{equation}
where $\mW^g$ denotes the $\mW$ that is obtained by the action of the group element $g$
as described by equation (\ref{eq:orbactionz}). Notice that in the generic case we are describing, all of these kets are linearly independent, because their eigenvalues with respect to the commuting $\mW$ are different for at least one such variable.

Starting from a single ket $\ket w$, we find that the action of the group generates images of $\ket w$ characterized exactly by the label  of group elements $g$. Moreover, we find that the action of the group on this basis is by permutations that exactly follow the group multiplication. This is, the typical irreducible (generated by $\ket w$) can be labeled by the group elements $g\in \Gamma$, and the action of $\Gamma$ on these states is the same action of $\Gamma$ on $\Gamma$ itself: by permutations. This representation of the discrete group algebra is the group algebra $\BC \Gamma$ itself as a left module over $\BC \Gamma$. This is called the regular representation of the group. If we decompose it into
irreducibles, we find always that
\begin{equation}
\BC \Gamma \simeq \sum_{i\in {irreps(\Gamma)}} {\dim(R_i)} R_i
\end{equation}
where $\mathbb{C}\Gamma$ contains each irreducible representation $R_i$ of $\Gamma$ $dim(R_i)$ times. This can be found in standard texts in representation theory of finite groups, (see \cite{FH}, p. 17 for example).

In the ABJM model, for each $\ket{w}$ with fixed eigenvalues under the $\mW$, we have a two dimensional space, characterized by $\ket{w,\pm}$, where $\pi^+\ket{w,+} =\ket{w,+}$, and
$\pi^+\ket{w,-}=0$. These are the two eigenspaces for the nodes of the ABJM  quiver.
The argument we did above works for each of these two eigenspaces. Also notice that it was not particularly important which group $\Gamma$ we used. When we look at this information and compare to the quiver diagrams presented in section \ref{sec:theories}, we see that a
bulk representation (generic) has ranks $\dim(R_i)$ on each of the $\pm$ nodes. Remember that these also have levels proportional to $\dim(R_i)$.

Each of these eigenspaces of the $\mW$ would be considered as a brane in the ABJM or KW theory, where the brane positions are inferred from the $w^{i,j}$ eigenvalues. What we see is that we have produced brane positions and their images in the conifold geometry. This way of proceeding makes it clear that we can analyze the theory with algebraic methods in a way that parallels very closely our geometric thinking on orbifold spaces.

So far, we have only solved the gauge invariant holomorphic data for a single brane. This would be enough to characterize the moduli space in four dimensional gauge theories. However, the three dimensional case is more involved, as some of the equations require real variables and the full $\BC^*$ algebra structure.

Notice now that if we consider the $\mU, \mV$ variables, they have all the same eigenvalues for all the $\ket {w,\pm}$. Thus, on each of these solutions, these variables belong to the center of the algebra. After all, they are proportional to the identity in any irreducible representation. We have not show that these are irreducibles of the full $\BC^*$ algebra, but they are irreducibles of the solutions of the  $F$-term equations.

 Let us show that given this Hilbert space associated to the regular representation, we can completely solve for the set of representations. This is, given the $\mW,\mU,\mV$ as scalar values, satisfying the relations we have described, we want to solve for $A,B$ variables.

By direct evaluation, we can find that $\mW^{1,2}= a^1_1b^2_1= a_2^1 b_2^2 \dots  $. We can choose the orbifold algebra to act also by leaving $A^1,B^2$ invariant (this corresponds to choosing phases so that $\phi$ cancels the phase of $\eta^{1/2}$). Denote the basis of the regular representation by $| w\eta^{j}\pm\rangle$, where $ w\eta^{j}$ is the eigenvalue of $\mW^{1.1}$. Then in this basis
\begin{equation}
A^{1}| w\eta^{j}+\rangle=a_{j}^{1}| w\eta^{j}-\rangle
\end{equation}
since we choose $\Gamma$ to act trivially on $A^{1}$
\begin{equation}
(A^{1})^{g}| w\eta^{j}+\rangle=a_{j}^{1}| w\eta^{j}-\rangle=a_{j-1}^{1}| w\eta^{j}-\rangle
\end{equation}
Thus, the $a_i^1$ must be equal to each other, as well as the $b_i^2$. This reflects the fact that we can do this operation also by a gauge transformation.

Given this information, we learn that $\mW^{1,1}= \diag( a^1 b_i^1)$, and that the action of the group on the $\mW^{1,1}$ forces the $b_j^1 = b_0^1 \eta^{-j}$, so that they are all the same up to a phase. Remember that
once we make a gauge choice for $A$, there is no more freedom on the $B$. So, in matrix notation, we have that $A$ is block diagonal
\begin{equation}
A^1=  \begin{pmatrix} \begin{pmatrix} 0& a^1\\
0&0\end{pmatrix} & 0 & \dots\\
0&  \begin{pmatrix} 0& a^1\\
0&0\end{pmatrix} &\vdots\\
\vdots &\ddots &\ddots
\end{pmatrix} = a^1 {\bf 1} \otimes \begin{pmatrix} 0&1\\
0&0\end{pmatrix}
\end{equation}
and
\begin{equation}
B^2= b^2 {\bf 1} \otimes \begin{pmatrix} 0&0\\
1&0\end{pmatrix}
\end{equation}

We also get that
\begin{equation}
A^2 = a^2 \diag( \eta^j) \otimes \begin{pmatrix} 0&1\\
0&0\end{pmatrix}\quad B^1 = b^1\diag(  \eta^{-j})\otimes \begin{pmatrix} 0&0\\
1&0\end{pmatrix}
\end{equation}
These are tensor products of the 'Clock' matrix, times a matrix in the $\ket\pm$ basis.

 Similarly, the group generator acts as
 \begin{equation}
 e_g = \begin{pmatrix} 0& 0& \dots &1\\
 1&0&\dots &0\\
 0&1& \ddots &\vdots\\
 \vdots &\ddots &\ddots
 \end{pmatrix} \otimes \begin{pmatrix} 1&0\\0&1\end{pmatrix}
 \end{equation}
 so it is a tensor product of a 'Shift' matrix and the identity.

 If we now choose a different basis, we can diagonalize $e_g$ into a clock matrix, and then $A^2, B^1$ become shift matrices, giving us the usual quiver representation, where the eigenstates of $e_g$ are the vector spaces associated to the nodes of quiver diagram. This change of basis is a discrete Fourier transform\footnote{Indeed if we consider the quiver algebra spanned by $\xi,\omega$ and solve for representations of it, imposing the D-term constraints ($\Sigma\sim 1$), we arrive at the basis where the $e_{g}$ are diagonal.}.

 It is easy to check that these matrices suffice to reconstruct the $\mW^{1,2}, \mW^{2,1}$, and that $\mU,\mV$ can be computed easily. They all satisfy the relations that are needed, so this gives a representation of the holomorphic part of the algebra.

 In general supersymmetric theories we would expect the gauge group to be complexified in the superfield formulation. However, this is usually fixed by imposing the D-term constraints.  In our setup, we find that the equations that replace the D-terms are those that state that the auxiliary field of the gauge potential is composite
 \begin{equation}
 \Sigma \sim A^\dagger A+A A^\dagger -B^\dagger B-B B^\dagger
 \end{equation}
Explicitly, we have that $\Sigma$ is the set of usual D-terms of the field theory in four dimensions. Since $\Sigma$ commutes with $A,B$ and it is real, it is proportional to the identity. Hence, all the D-terms for the $(U(1)\times U(1))^N$ are the same: the only freedom we have is in changing the scale of $a^s,b^s$ by a complexified gauge transformation, but we can not do that independently at each node, because that would modify the D-terms and we would not be able to satisfy $[\Sigma,B]=[\Sigma,A]=0$.

Let us see how this argument would work in more detail. If we consider a four dimensional theory, like the Klebanov-Witten conifold theory for the $U(1)\times U(1)$ gauge group,
the complexified gauge transformations can act on the fields as follows $A^{1,2}\to \exp(\gamma) A^{1,2}$ and $
B^{1,2}\to \exp(-\gamma) B^{1,2}$. The D-term equations of motion, that are given by
\begin{equation}
 |A|^2-|B|^2 -\alpha=0
\end{equation}
where $\alpha$ is a FI term that we fix to some value, need to vanish. Under these gauge transformations we can set  reference values where $|A_0|^2=|B_0|^2= \Omega/2$. We can then solve the D-term equations, by using a $\gamma$ such that
\begin{equation}
\Omega\sinh(2\gamma)= \alpha
\end{equation}
and since the function $\sinh$ covers the real line, there always exists a solution to this equation. This gives us a parametrization of the moduli space.

Here, we find that in the case of three dimensional theories we are studying the FI terms must all be essentially equal (the Chern-Simons level of the node appears as part of the calculation). For each set of values of the $\alpha$ parameter, there is a unique set of real exponents that solves the corresponding set of equations. However, these are the complexified gauge transformations that commute with the action of $\Gamma$, so they are diagonal in the basis where the action of $\Gamma$ has been diagonalized. This is different than the basis we chose above where the $\mW$ are all diagonal. We still will use the same letters to label the representations, with the understanding that there is a linear transformation between the $a^{1,2}$, and a discrete Fourier transform $\tilde a^{1,2}$, and $b^{1,2}$ gets also replaced by  $\tilde b^{1,2}$. In the quiver, these are the variables $Z$ and $W$. The D-term equations are given by
\begin{equation}
\sigma_{2l}=\frac{1}{4K}\Big[Z_{2l-1}^{\dag}Z_{2l-1}-W_{2l-1}W_{2l-1}^{\dag}+Z_{2l}^{\dag}Z_{2l}-W_{2l}W_{2l}^{\dag}\Big]
\end{equation}
and similar for $\sigma_{2l+1}$. Given that the $\sigma$ must all be equal to each other in an irreducible representation
(this is an application of Schur's lemma, since $\Sigma$ commutes with everything), there is only one degree of freedom
to tune, that is the value of $\sigma$ itself, on any one node.

So if the $|A|^2$ and $|B|^2$ are independent of the nodes that we are considering, we obviously solve these equations, and this solution is unique for each $\sigma$. There is still one real parameter $\gamma$ that can be used on all $A$, $B$ with the same weights as the conifold which we are free to vary (the diagonal $U(1)$ gauge transformation). This parameter, and its corresponding complexified phase give us that the set of representations of the $\BC^*$ algebra is one complex dimension higher than the same set of solutions in the four dimensional theory, for all setups that correspond to a single brane. We explained this in section \ref{sec:modpspace}. This phase is a gauge redundancy at the level of these fields. However, we know that moduli spaces should be complex, so $\sigma$ will end up complexified in the true moduli space. This has it's source in the dual scalar of the photon, whose vev is not manifestly present in the Chern-Simons lagrangian formulation. However, just keeping this phase can account for it. In a $U(1)$ theory we fix the gauge of that dual scalar to a fixed value, leaving only discrete gauge transformations that keep that dual scalar fixed.

Thus, the general brane is described by four complex numbers $a^1, a^2, b^1, b^2$. There is one phase redundancy of gauge transformations. However, when we include the Chern-Simons degrees of freedom, this becomes a discrete phase rather than a continuous one.

Notice also that if we fix $a^1$, in the representation there is a discrete identification
\begin{equation}
(a^1,a^2,b^1,b^2) \sim (\eta a^1,a^2,b^1,\eta^{-1}b^2)\label{eq:discreteid}
\end{equation}
 after a simple change of basis. We had a representation classified by these up to the (cyclic) discrete permutations of the eigen-blocks of the $A,B$ matrices that keep $e_g$ invariant. This discrete identification is the fact that the discrete symmetry of the original quiver was gauged, so that we can not tell apart a brane from its image.

So we have shown that the method of images lets us construct a solution of the equations in the quotient theory by the method of images. That solution, for a single brane in the bulk and its images, is an irreducible representation of the algebra. The non-degeneracy of the eigenvalues of $\mW^{i,j}$ guarantee this.

\subsubsection{The singularities and fractional branes}

The next step is to analyze what happens at the non-generic points of the 'orbifold of the conifold'. These are the locus where the $\mW^{i,j}$ degenerate. This is a set of positions where the $\mW^{i,j}$ are repeated between a brane and its image. Such degeneracy implies that there is a $1\neq g\in \Gamma$ that does not change the position of the brane in the conifold. Indeed, it is a subgroup of $\Gamma$ that has this property, and the fixed point is an orbifold singularity.
If at least one of the $\mW^{i,j}$ is non-zero, then this is not at the tip of the cone, and we would expect locally that we have a
curve of such singularities, because the geometry is a cone and the group identifications are compatible with rescalings in the cone. For the Calabi-Yau threefold,  this corresponds locally to a $\BC^2/\tilde \Gamma \times \BC^*$ singularity. These are generally classified by $ADE$ groups. In this particular case, we get an $A_{n-1}$ singularity.  The general wisdom is that a brane hitting such a singularity will split according to the irreducible representations of $\tilde\Gamma \subset \Gamma$. This is easy to see. We started with the
regular representation of $\Gamma$. When we reach the singularity, the degeneracy of the subspaces are classified by the regular representation of $\tilde \Gamma$. This is because these subspaces are classified by the eigenvalues of the $\mW$. The regular representation of $\tilde \Gamma$ splits these eigenspaces into irreducible representations of $\tilde \Gamma$. One can show that these do not mix when we take into account the rest of the group elements not in $\Gamma$, because the group $\Gamma\times \tilde \Gamma$ acts naturally on $\Gamma$ by a left of action of $\Gamma$, and a right action of $\tilde \Gamma$. This guarantees that the actions can be made compatible. For the abelian group we are considering it is always obvious.

The location of the singularities of the Calabi-yau three-fold can be understood by noticing that these are singularities
of the equations defining the Calabi-Yau geometry. These are given by the locus
\begin{equation}
\mU= \mV=0\quad
\mW^{1,2} \mW^{2,1}=0
\end{equation}
There are two such lines of singularities in our case. Those where $\mW^{1,2}=0$, or those where $\mW^{2,1}=0$.
Let us consider the second such locus first $\mW^{2,1}=0=\mU=0=\mV=0$.
Having these set equal to zero gives us the following locations in terms of the $a^1,a^2, b^1, b^2$ variables
\begin{eqnarray}
a^1 b^1= a^2b^2= a^2 b^1= 0\\
a^1 b^2\neq 0
\end{eqnarray}
This makes us set $a^2=0=b^1$. From the action given in equations (\ref{eq:orb2}), these are exactly the locus where the
action on the fields has a fixed point.

For the other singularity, we would have $a^1=0, b^2=0$ and we would at first think that equation (\ref{eq:orb2}) would imply that it is not a fixed point. However, another gauge transformation is possible that will let us keep $A^2, B^1$ fixed, while transforming $A^1, B^2$ with phases. With respect to this action on the fields, the fields are a fixed point of the orbifold group. This shows why it is so important to keep track of the gauge redundancy when deciding if we have a fixed point of the orbifold action or not.

These solutions with singularities give us a copy of $\BC^2$ for each fractional brane at a singularity.
If we only keep the fields that are non-zero in the quiver, we see that we get a picture as shown in figure \ref{fig:QuiverBK2.eps}
\myfig{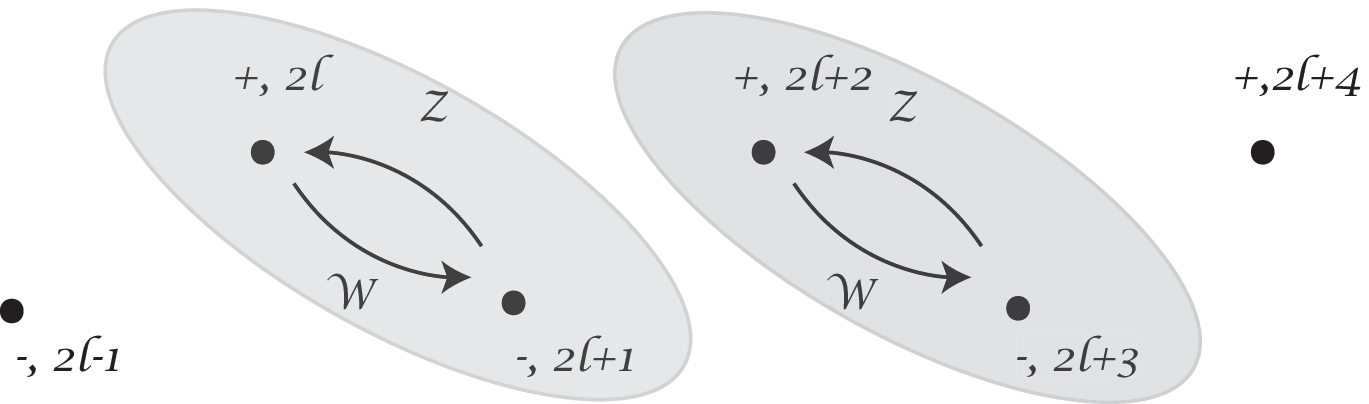}{6}{Quiver with groups of nodes shaded according to vevs}

There are $n$ such nodes. The fact that some nodes are not connected, means that the representation is reducible: we can not go from the vector spaces $\CV$ joining them by a non-zero map. The set of $n$ irreducibles is exactly what we expect from brane fractionation at $A_{n-1}$ singularities \cite{DM}: there should be $n$ irreducibles, one for each blow-up cycle, and one extra for the extended root of the affine Dynkin diagram of the $A_{n-1}$ system.

The arrows that are missing are the fields that can become massless when two of these fractional branes are in the same position in the CY-geometry. If we contract the shaded areas to points,  these missing
arrows would give us the quiver of the $A_{n-1}$ singularity.

Each of those blocks would correspond to a branch of moduli space for a $U(1)\times U(1)$ subquiver\footnote{An equivalent way to see this is by looking at the simple modules of $\mathcal{A}\boxtimes \mathbb{Z}_{n}$ at the singular points of the moduli space. In these points we get a collection of irreducibles that are exactly the fractional branes, $\lim_{a^{2},b^{1}\rightarrow 0}R(a^{1},a^{2},b^{1},b^{2})=\bigoplus_{l=1}^{n}R_{l}(a^{1},b^{2})$.}. Moreover, we can see easily that the vevs of the $A,B$ fields can be complete uncorrelated between the different fractional branes. Thus, there is no unique $a^1, b^2$ characterizing them, instead there is one such value for each subquiver.

For the other set of singularities, the splitting is different, along the other diagonal  cells in the quiver graph.

There is an extra potential set of singular solutions where $\mU=\mV=\mW^{1,2}=\mW^{2,1}=0$, which are characterized by either $a^1, a^2=0$, or by $b^1, b^2=0$. These solutions in the four dimensional Klebanov-Witten theory would correspond to vacua at the tip of the cone, unless the FI-terms are set to be different from zero. In such a case, these would give rise to points in the exceptional divisor of the blow-up of the conifold. The coordinates $(a^1, a^2)$ (or $(b^1,b^2)$ for a different choice of the FI-term)  would be the homogeneous coordinates on this $\CP^1$. Generically,  these are not fixed points of the orbifold group if $a^1$ and $a^2$ are different from zero.
In the $\BC^*$ algebra setup, we see that the non-holomorphic coordinates $a^1 a^{2*}$ would be invariant under the gauge transformations, but would transform, hence these are not fixed points in the blow-up.
 And if one of them is zero, it is in the locus that the subquivers described above cover. These singularities do not intersect in the blow-up, so there is no additional fractionation.

This shows us a nice correspondence of the singularity structure of the Calabi-Yau geometry associated to the four dimensional theory relative to the singularities of the moduli space of the three dimensional theory.  The two lines of $A_{n-1}$ singularities in the Calabi-yau geometry become two copies of $\BC^2$, with the same $A_{n-1}$ singularity
around them. In this case, these are all the singularities of the geometry that are not the tip of the 4-dimensional complex geometry.

\subsection{The complete moduli space}

We have described how to build some solutions of the vacuum constraints of the theory. For theories that have
couplings of single trace type, there is a general recipe to build the moduli space from the components we have studied so far \cite{BJL}. The idea is that solutions of the vacuum equations for block diagonal matrices
can be obtained if every block satisfies these equations on its own.

We have classified the solutions in terms of representations of the quiver algebra with vacuum relations. Let us call them $R_\alpha$. The $\alpha$ are the parameters that describe a particular irreducible representation of the quiver algebra (for example, $R(a^1, a^2, b^1, b^2)$, or $R_{l}(\hat a^1, \hat b^2)$ for the fractional branes), and they also contain discrete labels for the fractional branes: the sublock of  nodes of the quiver that the representation covers.  We have also kept one extra degree of freedom for each brane that arises from dual photons on the theory. This vev does not have any implications at the level of perturbation theory: for example, masses of off-diagonal fields are independent of these phases, as can be seen very explicitly in the ABJM model \cite{BT}. Here, the same equations work by the method of images, as expected from general features of the construction \cite{DM}. Given these blocks, one can build new solutions of the equations of the vacuum by taking direct sums:

\begin{equation}
R = \oplus_{\alpha} R_\alpha
\end{equation}

This general solution by a representation of the algebra solves all the equations of motion of the vacuum. The representation space has constraints from the ranks of the gauge groups to be fixed, but are otherwise unconstrained.

For each brane there is a massless photon, and the dual scalar action can be used in the low energy effective action. This is allowed since in the generic representation of this set there are no massless charged particles. Thus, we need to remember that for each brane there is a circle direction that is invisible in perturbation theory.

 The labels $\alpha$ can vary for each brane, so the moduli space generically described a collection of branes at various loci.
 The sum is unordered, because how to organize blocks into matrices is a gauge choice. Thus, the general moduli space is a generalization of a symmetric product space, and give an appropriate notion of a symmetric product for a non-commutative geometry \cite{BJL}. It would be a standard symmetric product if all branes could be exchanged with each other by motions in parameter space. However, the process of brane fractionation involves processes where one brane can split into many. These give rise to different branches in moduli space. The simplest example of a variety with two
 branches meeting at the origin is the subset of $\BC^2$ characterized by the equation $xy=0$. There are two branches, $x=0$ and $y=0$, each of them a complex  line. These two meet at the origin.  A general system of branes where branes can fractionate and give rise to new branches of moduli space has a similar structure. This implies that in the chiral ring there will relations like the one above, $xy=0$, where $x$ and $y$ can be elements of the chiral ring, but not their product.

 These relations become rather complicated for the chiral ring of theories with many branes. But if we know what the geometry of the moduli space looks like, then the relations are implicit in the geometry. We will not address this issue further in this paper.

Also, for each brane there are discrete phases that need to be taken into account. These give identifications between the $R_\alpha$ parameters that we need to analyze further. These can be conveniently described in terms of the chiral ring elements. We will describe these in what follows.

\subsection{The chiral ring}

As described previously, the chiral ring can be obtained from a semiclassical quantization of solutions of the BPS equations on the sphere (we quantize the space of those solutions by wiring a wave function on them and counting the allowed wave-functions). We will describe the chiral ring here in this manner, rather than as local words on the elementary fields. The BPS equations force the fields to be spherically symmetric and to evolve according to their R-charge.

  Moreover, we saw that the classical solutions require that the scalar field expectation values are in the moduli space of the theory in flat space. The semiclassical quantization will place constraints that will determine the full topology of the moduli space of vacua in the end.  What we have calculated so far is a cover of the moduli space of vacua, as there
are possible identifications between configurations that we have not described yet. We have already constructed the full basic structure of moduli space. Since the moduli space of vacua has different branches, we need to analyze these equations in different branches to obtain results.

The next step is to include the equations of motion of the Chern-Simons gauge fields and to perform the correct holomorphic quantization of moduli space. In particular, we have found the wave functions on moduli space are naturally holomorphic. So the chiral ring is identified exactly with holomorphic wave functions on moduli space.

There is one last thing to consider. That is that the moduli space is a generalization of a symmetric product, which is a collection of representations with various charges assigned to them: fractional branes have additional discrete charges. These are counted by the rank of the different gauge groups.

This means that wave functions need to be symmetrized between components. This symmetrization will be assumed throughout. It gives rise to a natural structure in terms of products of traces (summing over branes). This is automatically invariant when we permute branes. Thus, we can analyze the chiral ring one brane at time and this description is sufficient for describing the whole chiral ring.

We will do this in what follows. First we need to verify what the classical equations of motion say about the chiral ring classical BPS states.

When we consider the theory with the fields  in $S^{2}\times \mathbb{R}$ we get a coupling of the background curvature to the scalars. Since on BPS configuration the potential vanishes, the effective action on these reduced configurations can be without potential terms. Moreover, if we apply this to our case, ignoring fermionic and potential terms the effective action on BPS states takes the form
\begin{eqnarray}
\mathcal{S}&=&-\frac{\kappa}{4\pi}\int d\Omega d\tau\sum_{l=1}^{2n}(-1)^{l}\epsilon^{\mu\nu\lambda}Tr\Big(A^{(l)}_{\mu}\partial_{\nu}A^{(l)}_{\lambda}+\frac{2i}{3}A^{(l)}_{\mu}A^{(l)}_{\nu}A^{(l)}_{\lambda}\Big)\nonumber\\
&-&\int d\Omega d\tau\sum_{l=1}^{2n}Tr\Big((\mathcal{D}^{\mu}Z_{l})^{\dag}\mathcal{D}_{\mu}Z_{l}+(\mathcal{D}^{\mu}W_{l})^{\dag}\mathcal{D}_{\mu}W_{l}+\frac{1}{4}W^{\dag}_{l}W_{l}+\frac{1}{4}Z^{\dag}_{l}Z_{l}\Big)
\end{eqnarray}
where
\begin{eqnarray}
\mathcal{D}_{\mu}Z_{2l}&=&\nabla_{\mu}Z_{2l}+iA^{(2l+1)}_{\mu}Z_{2l}-iZ_{2l}A_{\mu}^{(2l)},\nonumber\\
\mathcal{D}_{\mu}Z_{2l-1}&=&\nabla_{\mu}Z_{2l-1}+iA^{(2l-11)}_{\mu}Z_{2l-1}-iZ_{2l-1}A_{\mu}^{(2l)},\nonumber\\
\mathcal{D}_{\mu}W_{2l}&=&\nabla_{\mu}W_{2l}+iA^{(2l)}_{\mu}W_{2l}-iW_{2l}A_{\mu}^{(2l+1)},\nonumber\\
\mathcal{D}_{\mu}W_{2l-1}&=&\nabla_{\mu}W_{2l-1}+iA^{(2l)}_{\mu}W_{2l-1}-iW_{2l-1}A_{\mu}^{(2l-1)},
\end{eqnarray}
for spherically symmetric configurations
\begin{eqnarray}
\mathcal{D}_{i}F^{(l)}_{\mu\nu}=0\qquad\nabla_{i}Z_{l}=\nabla_{i}W_{l}=0\qquad\forall l\qquad i=\varphi,\theta
\end{eqnarray}
then $F^{(l)}_{0i}=0$ and by gauge fixing $A^{(l)}_{0}=0$, and $F^{(l)}_{\varphi\theta}=\tilde{\Phi}^{(l)}$, where $\tilde{\Phi}^{(l)}$ is a diagonal constant matrix. The magnetic fluxes $\Phi^{(l)}=\int_{S^{2}}\frac{\tilde{\Phi}^{(l)}}{\sin(\theta)}$ are classically quantized.

For a single brane in the bulk, the BPS equations mandate that
$\mathcal{D}_{\mu}Z_{l}=\mathcal{D}_{\mu}W_{l}=0$ for all $l$, and so $\tilde{\Phi}^{(l)}=\tilde{\Phi}$ for all $l$. If the fluxes would not be the same the matter would be charged under a magnetic monopole background and it would not be spherically symmetric (monopole spherical harmonics carry spin). This is the same reasoning found in \cite{BT,BP}.

The equation of motion for $A^{(l)}_{\mu}$ vanishes identically for $\mu=\theta,\varphi$, the e.o.m for $\mu=0$ gives
\begin{eqnarray}
-\frac{\kappa}{\pi\sin(\theta)}F^{(2l)}_{\theta\varphi}=-i\dot{Z}_{2l}^{\dag}Z_{2l}+iW_{2l}\dot{W}^{\dag}_{2l}-i\dot{Z}^{\dag}_{2l-1}Z_{2l-1}+iW_{2l-1}\dot{W}_{2l-1}^{\dag}+h.c.,\nonumber\\
\frac{\kappa}{\pi\sin(\theta)}F^{(2l-1)}_{\theta\varphi}=i\dot{Z}_{2l-2}^{\dag}Z_{2l-2}-iW_{2l-2}\dot{W}^{\dag}_{2l-2}+i\dot{Z}^{\dag}_{2l-1}Z_{2l-1}-iW_{2l-1}\dot{W}_{2l-1}^{\dag}+h.c.,
\end{eqnarray}
In this prescription, the $Z$ and $W$ fields satisfy the equation of an harmonic oscillator $\dot{Z}=i\frac{1}{2}Z$, $\dot{W}=i\frac{1}{2}W$ as they are of dimension $\frac{1}{2}$. Then
\begin{eqnarray}
-\frac{\kappa}{\pi\sin(\theta)}\tilde{\Phi}=-|Z_{2l}|^{2}+|W_{2l}|^{2}-|Z_{2l-1}|^{2}+|W_{2l-1}|^{2},\nonumber\\
\frac{\kappa}{\pi\sin(\theta)}\tilde{\Phi}=|Z_{2l-2}|^{2}-|W_{2l-2}|^{2}+|Z_{2l-1}|^{2}-|W_{2l-1}|^{2}.
\end{eqnarray}
where $\kappa$ is the Chern-Simons level.

When we are away from the singularities, we can substitute the solutions of these equations for a brane in the bulk.
These are characterized by $a^1, a^2, b^1, b^2$, so we find that
\begin{equation}
\frac{\kappa}{\pi}\Phi= \int_{S^{2}} |b|^2-|a|^2 \label{eq:const}
\end{equation}
and in the Hamiltonian all of $a^{1,2}$, $b^{1,2}$ have the same frequency, $\omega= 1/2$.

The effective Hamiltonian for these variables is of the form
\begin{equation}
H_{eff}= ( i \Pi_a a + i \Pi_b  b)
\end{equation}
where $\Pi_a$ and $\Pi_b$ are the canonically conjugate momenta to the $a,b$ variables. This obviously reproduces the BPS equations of motion $\dot a = i a/2$, etc.

We also have to take into account the constraint (\ref{eq:const}).  Moreover, we have the identifications on the parameters $a$ that are characterized by the discrete action (\ref{eq:discreteid}), or some equivalent identification that depends on a gauge choice for one variable.

When replacing all the $Z,W$ by their expressions in terms of $a,b$, we find that the $R$-charge is given by
\begin{equation}
Q_R =  n  \left(  -i \dot {\bar a} a- i \dot {\bar b} b\right)
\end{equation}
and comparing with $H_{eff}$, we find that $\Pi_a = -in \dot{\bar a}$. This can be derived also by direct substitution in the original lagrangian. The factor of $n$ is here because we have to sum over all $Z,W$ identical factors.

The $a$ commute with each other and with $b$ on the set of BPS solutions, while their complex conjugate variables have non-trivial commutation relations with $a,b$ on the reduced phase space of solutions (see \cite{BP} for more details).
A holomorphic quantization will give us polynomials in the $a$ variables, while the canonical conjugate momenta get represented by derivatives $\Pi_a\sim i\partial_a$.

As can be seen, the effective Hamiltonian is the same as that for a Harmonic oscillator in four dimensions (four complex dimensions since we are on phase space), and the natural variables are holomorphic. Thus, wave functions are polynomials in the $a,b$, and the energy of a monomial is the degree of the monomial divided by two. A typical wave function will be as follows
\begin{equation}
\psi \sim (a^1)^{k_1} (a^2)^{k_2} (b^1)^{m_1} (b^2)^{m_2}
\end{equation}
Since the system has an extra $U(1)^3 $ symmetry, we can choose wave functions that are eigenfunctions of these $U(1)$ charges (they count the number of $a^{1,2}, b^{1,2}$) and these are just monomials.
However, not all of these are allowed. There are constraints that need to be satisfied.

First, for the standard integral quantization of the magnetic flux requires that $\int_{S^2} \frac{\tilde{\Phi}}{\sin(\theta)}= 2\pi m$, from which
\begin{equation}
\kappa m = \frac 1n (k_1+k_2-m_1-m_2)
\end{equation}
so that $k_1+k_2-m_1-m_2$ is a multiple of $\kappa n$.  This combines the classical integrality of the magnetic flux with the integral quantization of harmonic oscillator wave functions. As shown in \cite{BP}, one can also have fractional flux on all fluxes simultaneously. This enlarges slightly the space of possibilities. The detailed study of the fractional flux configurations is beyond the scope of the present paper, suffice to say, these are described by D4-branes in the dual theory and gives rise to a discrete fibration over a symmetric product of the orbifolded space.

Similarly, the other identification from redundancy of the representation content should impose that the wave function is single valued under those replacements (the same prescription was used in \cite{Bcor,Bcot}). This forces us to have $m_2-k_1$ being a multiple of $n$. Thus, the allowed polynomials are those that correspond to the invariant ring of $\BC^4$ under the following two actions
\begin{eqnarray}
(a^1, a^2, b^1, b^2) &\to& ( \beta a^1, \beta a^2, \beta^{-1} b^1, \beta^{-1} b^2)\\
(a^1, a^2, b^1, b^2) &\to& (\eta a^1,  a^2,  b^1,\eta^{-1} b^2)
\end{eqnarray}
where $\beta^{n\kappa}=\eta^n=1$ are primitive roots of unity.

This is, we describe this way a single brane on the quotient space $\BC^4/ \BZ_{\kappa n} \times \BZ_n$.
When we have many representations, the flux quantization is done on each of them independently. This result was obtained first in \cite{TY,IK}.

For the case $m=1$, the minimal energy solutions have $k_1+k_2=\kappa n$ and $m_1=m_2=0$. The energy of this state is $\kappa n$. Also, $k_1-k_2$ should be a multiple of $n$. The simplest solution has $k_1=\kappa n$, $k_2=0$.

This covers the chiral ring elements that on the cylinder probe the 'brane in the bulk' solutions. There are also the
chiral ring elements that probe the fractional brane branches. These are classical solutions in a subquiver with group $U(1)\times U(1)$, and can  also be analyzed easily.

In this subquiver, we have variables $\tilde a_i^1, \tilde b_i^2$ for only one $i\neq 0$. Again, the effective Hamiltonian will be that for a harmonic oscillator in two dimensional phase space (we only have two coordinates).

The flux quantization condition becomes
\begin{equation}
\kappa m = n_1-m_2
\end{equation}
which shows that the allowed monomials are given by $(\tilde a^1_i) ^{n_1} (\tilde b^2_i)^{n_2}$, and that $n_1-m_2$ are a multiple of $\kappa$, the level. This is the same reasoning for all the possible fractional brane representations parametrized by $i$. There is a similar set of states for the other singularities. Again, the simplest operator with non-zero flux appears for $m=1$, and $n_1=\kappa$, while $m_2=0$.

This means that the fractional brane branch corresponds to the invariant ring of a $\BC^2/\BZ_{\kappa}$ quotient, without a factor of $\kappa n$ appearing in it. Notice that the only states that can sense the multiples of $\kappa$ have flux and are therefore associated to monopole operators.

When this fractional branch is considered, there is monodromy along the circle fiber with respect to the branes in the bulk, since in the bulk the circle fiber is divided by
$\kappa n$, rather than just $\kappa$. This monodromy is similar to fractional branes in orbifolds with discrete torsion \cite{BLdt} where a similar monodromy of fractional branes is encountered, except that in that case the monodromy is already visible with perturbative gauge invariant operators and does not require monopole operators.

\subsection{Matching monopole states to the AdS dual states}

We have found quite a variety of solutions of the equations of motion of the fields on a sphere that we can identify with states on the cylinder under quantization. We want to compare these states to those that are expected from the $AdS$ dual theory on $AdS_4\times S^7/\BZ_{\kappa n}\times \BZ_n$, where the $\BZ_{\kappa n}$ acts along the Hopf fibration, and the corresponding type IIA theory on
$AdS_4\times \CP^3/\BZ_n$ with fluxes.  Here, the $\CP^3$ is the base of the Hopf fibration of $S^7$. The fiber of the Hopf-fibration is the circle of gauge transformations that the dual photon makes physical. Hence, from the point of view of the natural fields of the quiver, it is very closely related to gauge transformations (see also \cite{BT}).

In terms of the $a^1, a^2, b^1, b^2$ coordinates that we have been describing so far, the homogeneous coordinate ring of $\CP^3$ are formed by $a^1, a^2, (b^1)^*, (b^2)^*$, similar to how the $\CP^3$ coordinates of the ABJM model work. Notice that in the ABJM model the matter fields associated to $A, B^*$ have the same gauge theory representation content. Hence they can be grouped together, and their ratios can be considered to be gauge independent.

In the type IIA picture, the operators that carry momentum along the Hopf fiber have D0-brane charge. The coordinates $a^{1,2}, (b^{1,2})^*$ carry positive charge, since they have period one on the Hopf-fiber of $\CP^3$.

Also, for BPS states that are BPS with respect to our choice of ${\cal N}=2$ supercharge, $(b^{1,2})^*$ have the opposite time dependence than $a^{1,2}$. Thus, when we time evolve the system, the homogenous coordinates on $\CP^3$ change unless either $a=0$ or $b=0$.

We find that therefore the D0 brane charge should basically count the number of $a$ minus the number of $b$ letters in a monomial. Such monomials require magnetic flux. We therefore have to identify the uniform magnetic fluxes on the cylinder theory to give rise to the D0 brane charge.

Given the topological classification of line bundles on a sphere, for each node in the quiver diagram there is a determinant line bundle associated to it. The first Chern class of that line bundle is the sum of the fluxes on each of the eigenvalues.
This is not allowed to change, as it is an invariant under homotopies. Therefore we find that the total magnetic flux on each node is a topological invariant. These should be associated to conserved charges in the theory. Because all of these fluxes add to the notion of D0 brane charge, we should identify magnetic fluxes with brane charges.

Now, let us look at the tension of a D0 brane on $AdS_4\times S^7/\BZ_{\kappa n} \times \BZ_n$. A D0 brane can move along the Hopf fiber without any extra motion on the base, so from the point of view of $\CP^3$, it will stay at a fixed position. Since the Hopf fiber is reduced in size by a factor of $1/\kappa n$ relative to a natural $S^7$, the typical momentum along the fiber is $\kappa n$. This translates into an energy equal to $\kappa n/2$ in AdS units. This is exactly the energy of the simplest flux configuration we could find, associated to a brane in the bulk.

However, not all positions of a D0 brane in the bulk correspond to a holomorphic operator in the field theory. Only those where $b=0$ correspond to BPS states that saturate the correct BPS inequality. From the point of view of the type IIA theory, the D0 branes see a magnetic field on the $\CP^3$ base. There is a lowest Landau level associated to
these particles in a magnetic field. Since in this particular orbifold we preserve an $SO(4)$ R-charge, we find that the states in the lowest Landau level that we are describing have maximal angular momentum. This is why they reside in a submanifold of the set of possible D0-brane configurations.

If we excite some of the holomorphic $b$
monomials, we end up in a situation where the D0 brane moves. Again, this puts us in excited Landau levels, in the maximal angular momentum band. Again, these states are near the locus of the D0-branes that not move, until we go to very high excitations.

Notice that we can bring a BPS D0-brane near the orbifold singularities on $\CP^3$. At the singularity we expect the D0 branes to fractionate into $n$ fractional branes. Indeed, this is what we see. So we find that the total fluxes on the
nodes of the quiver must correspond to the number of fractional branes of each type.

Each of these should have a tension that is $1/n$ times the tension of D0 brane in the bulk. This is exactly what we find. The simplest fractional brane solutions carry $R$-charge equal to $\kappa$, rather than $\kappa n$. Since the $R$-charge is equal to the energy (tension) of the configuration, we find the expected result. Since
we have an $SO(4)$ R-charge, these charges necessarily are quantized in half-integer units. This is why the charge of a D0-brane in the bulk is $n$ times larger than a naive guess would have suggested: the fractional branes that combine to make it have charge that is a multiple of $\kappa$.  This explains why the orbifold where the branes move is in the end for branes on $\BC^4/\BZ_{\kappa n}\times \BZ_{n}$.

There are also solutions without magnetic flux that correspond to the 'fractional' brane classical solutions. These
just describe the expected massless modes arising from twisted sector strings at the singularity \cite{Gukov,BLdt}. Since these don't carry flux, they can be expressed as words in elementary fields: they are perturbative solutions of the theory with small energy.

We should also notice that the field theory permits us to have flux greater than one on a single eigenvalue for fractional brane solutions. These indicate a bound state of fractional branes at the singularity. Such states are usually forbidden for the theory on a flat space orbifold \cite{SetSt} (more information on the index computations that are necessary for these statements can be found in \cite{MNS}). Here, we see that these solutions can not be deformed by a small parameter into two fractional branes moving separately from each other (it requires a jump in flux from one eigenvalue to another). This suggest that these bound states are separated from the set of solutions with two fractional branes moving independently from each other by a potential barrier. Happily, fuzzy spheres in string theory can provide such a barrier in the presence of fluxes, as discovered by Myers \cite{Myers}. Indeed, this is how bound states of D0 branes are expected to be matched between the field theory solutions and the gravity dual \cite{S-JS}. Even though the configurations are abelian in field theory, they are non-abelian in the string dual.

\section{Aspects of non abelian orbifolds}

In this section we shall only sketch the results for other abelian and non-abelian orbifolds that can be obtained by following a similar path to the example wee have analyzed in detail.

Again, we can begin by analyzing the moduli space of the orbifolded theory by a group $\Gamma$. This is most  easily done by studying the method of images, and basically, we get that the moduli space of branes in the bulk should correspond to $N$ particles moving on $\BC^4/\BZ_{\kappa r} \times \Gamma$. One of the purposes of this section is to determine what the correct value of $r$ should be. For the $\BZ_n$ case studied above, we saw that $r=n=|\BZ_n|$. We will show that this generalizes to the order of the group.

To understand this, let us examine again the case $\Gamma=\hat D_k$ acting on one $SU(2)$ of the $SO(4)\simeq SU(2)\times SU(2)$-R charge that commutes with the choice of $\mathcal{N}=2$ superspace.

Such a $\Gamma \subset SU(2)$ would act only on the $B$ fields, lets say, but not on the $A$ fields. Again, the best way to describe the set of configurations is with the crossed product algebra. This is not identical to the quiver algebra any longer. Instead, they are Morita-equivalent. This means that the representation theory is the same, and it is parametrized by the same data. The crossed product algebra is the one that captures the method of images precisely.

This means that a brane in the bulk is again parametrized by four numbers $a^{1,2}, b^{1,2}$, with identifications on
the $a^{1,2}$ coordinates by $\Gamma$. The simplest singularities occur when $a=0$. These are fixed points under $\Gamma$. Branes fractionate at those locus. In the quiver diagram of figure \ref{fig:QuiverDn.eps} we turn off the crossed arrows, and we get a splitting of the diagram as follows in the figure\ref{fig:QuiverDnSplit.eps}.

\myfig{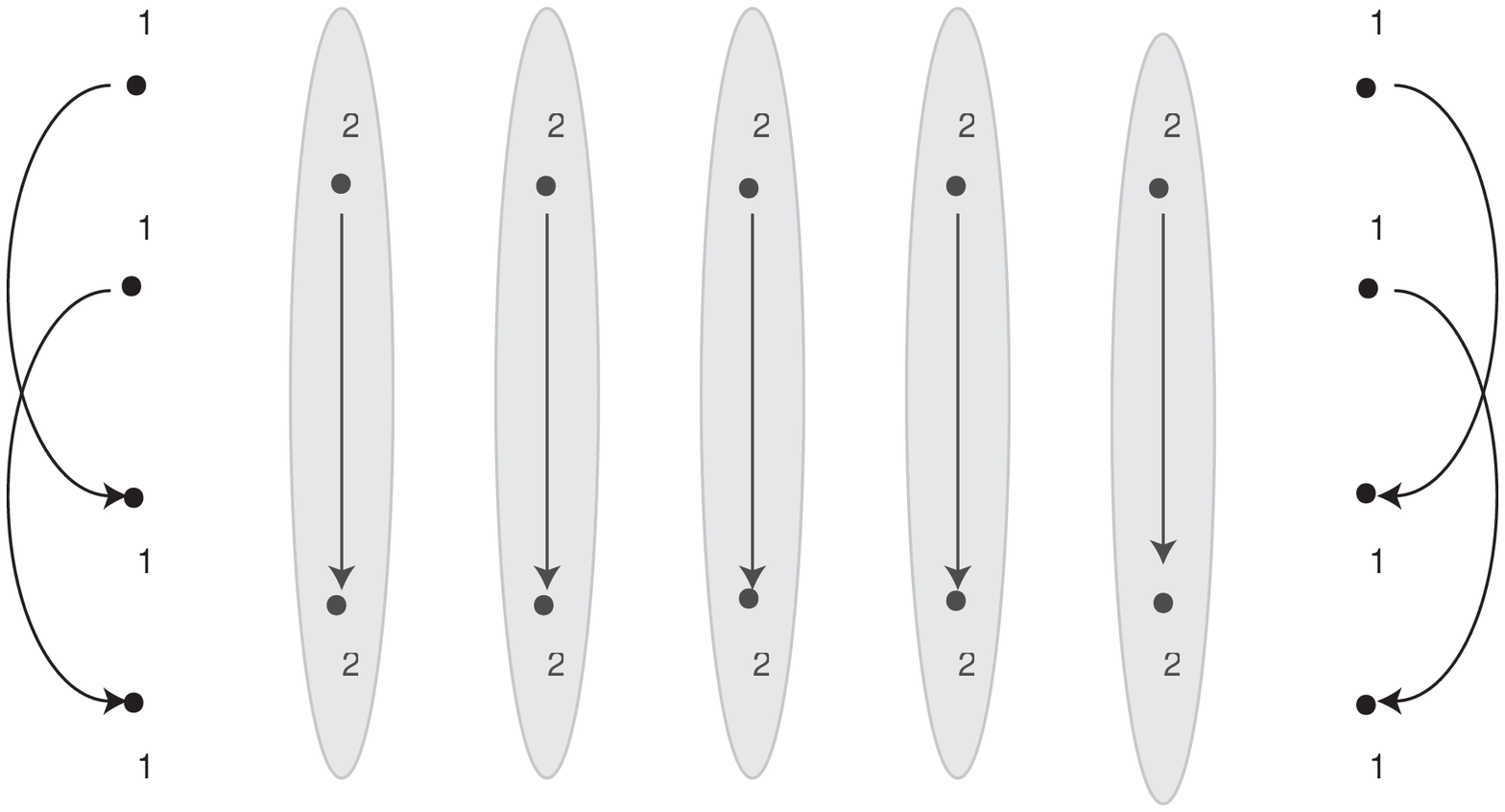}{8}{Splitting of the quiver diagram when the missing arrows are turned of to zero vev: the gray areas indicate contraction of nodes where the arrows that are turned on are connected. The curved arrows should be contracted similarly. The missing arrows would reconstruct a standard quotient for an ADE quiver. }

We should notice that again, the quiver splits into subquivers. Fractional branes and twisted sector states should be associated to these splittings. One of the interesting things that follows from the Douglas and Moore construction is that
fractional branes do not all have the same tension by the method of images. The tension of a brane associated to representation $R_i$ is proportional to $R_i$ \cite{DM} ( see also \cite{DDG}) . How does this get realized in the present context?

Again, flux on the subquivers should  be related to fractional brane charge. Remember that for the vevs to be spherically symmetric, the flux should be matched between nodes connected by arrows. This gives us freedom to have different fluxes on different subquivers thereby recovering fractional brane charges.

There is a new ingredient however. The level of the Chern-Simons fields on the nodes is proportional to the dimension of the representation $R_i$, times the basic level of the original ABJM model $\kappa$.
This means that when we match the minimal brane charge, associated to a node, the $R$ charge carried by the configuration is proportional to $\dim(R_i)$ (this is how the level of the Chern-Simons affects the equations of motion of the gauge fields and related it to the charge of the matter fields).

In equations
\begin{equation}
\kappa \dim (R_i) m = -m_1-m_2
\end{equation}

Notice that  in this equation $m_1,m_2$ are positive integers (this follows from chirality): there are no
chiral ring operators with $m=0, m>0$.
This would seem to imply that we can not have the opposite magnetic flux (corresponding to branes rather than anti-branes), because we only have positive contributions from matter to the charge. What we find instead is that states with opposite magnetic flux are necessarily anti-chiral (they require the $b$ to have the opposite time dependence).

Thus, the $R$-charge of these fractional branes is $\dim(R_i)\kappa/2$ and the D0 brane charge is $-\dim(R_i)/|\Gamma|$. By contrast, the $R$ charge of a D0-brane in the bulk is the sum of these R-charges with multiplicity and positive magnetic flux, giving us a tension of a D0-brane equal to
\begin{equation}
T_{D0}= \sum \frac{\dim(R_i)^2 \kappa}2 = \frac{|\Gamma|\kappa}2
\end{equation}
The equality of the sum of dimension squared of representations and the order of the group is a straightforward fact of discrete groups. It follows from the character of the identity of the regular representation. So from the fractional branes we constructed, it is simplest to build an anti-D0 brane with positive R-charge.

Given that the tension of this D0-brane dual object is exactly $|\Gamma|$ times larger than $\kappa$ shows that the circle fiber of the Hopf fibration is divided by a further factor of $|\Gamma|$ than that one provided by the $\kappa$ factor of the ABJM theory. This means that  these orbifold constructions should correspond to membrane theories on
\begin{equation}
\BC^4/ \BZ_{\kappa|\Gamma|}\times \Gamma
\end{equation}

This follows in the general case from solving the equations of motion with uniform flux on all the nodes for the regular representation: the flux quantization condition counts the number of $a$ minus $b$ fields at each node, but their normalization at each node differs (the fields are multiplied by Clebsch -Gordon coefficients after all). Also, the global normalization of the $a$, $b$ coordinates by the method of images is proportional to $|\Gamma|$, the number of copies
of a brane. These factors conspire to give us the above result in the general case.

Notice that there are other singularities of the group action in the type IIA picture (as would correspond to fixed points of subgroups of $\Gamma$ on $\CP^3$). These occur when the pair $(a^1, a^2)$ that can be used to describe a $\CP^2$, is at a fixed point of a subgroup of $\Gamma$ (they get multiplied by a common phase). At these singularities one can do a gauge transformation that keeps the pair fixed. This is a fixed point if $b=0$, as then the transformed configurations is equivalent to itself by a gauge transformation. These can be a $\BZ_{2n}$ singularity. It is easy to understand how the $\hat D_k$ quivers arise from orbifolding an $A_{2n-1}$ quiver \cite{BL} (see also \cite{BCD} for more related information about solving the equations for the matrix model realization of ADE quivers and related group theory constructions).

At these singularities, the $A$ fields can have vevs, but not the $B$ fields. The quiver splits differently, depending which $a$ is allowed to have a vev. A new ingredient is that the field $a$ can connect pairs of nodes with different level. The effective  $U(1)\times U(1)$ theory on a pair of nodes can not solve the equations of motion of the gauge fields.

There are new collections of fluxes that seem to work. These are given in the $\hat D_k$ case by branes with a $U(1)\times U(1)\times U(1)$ theory, as shows by the following figure.
\myfig{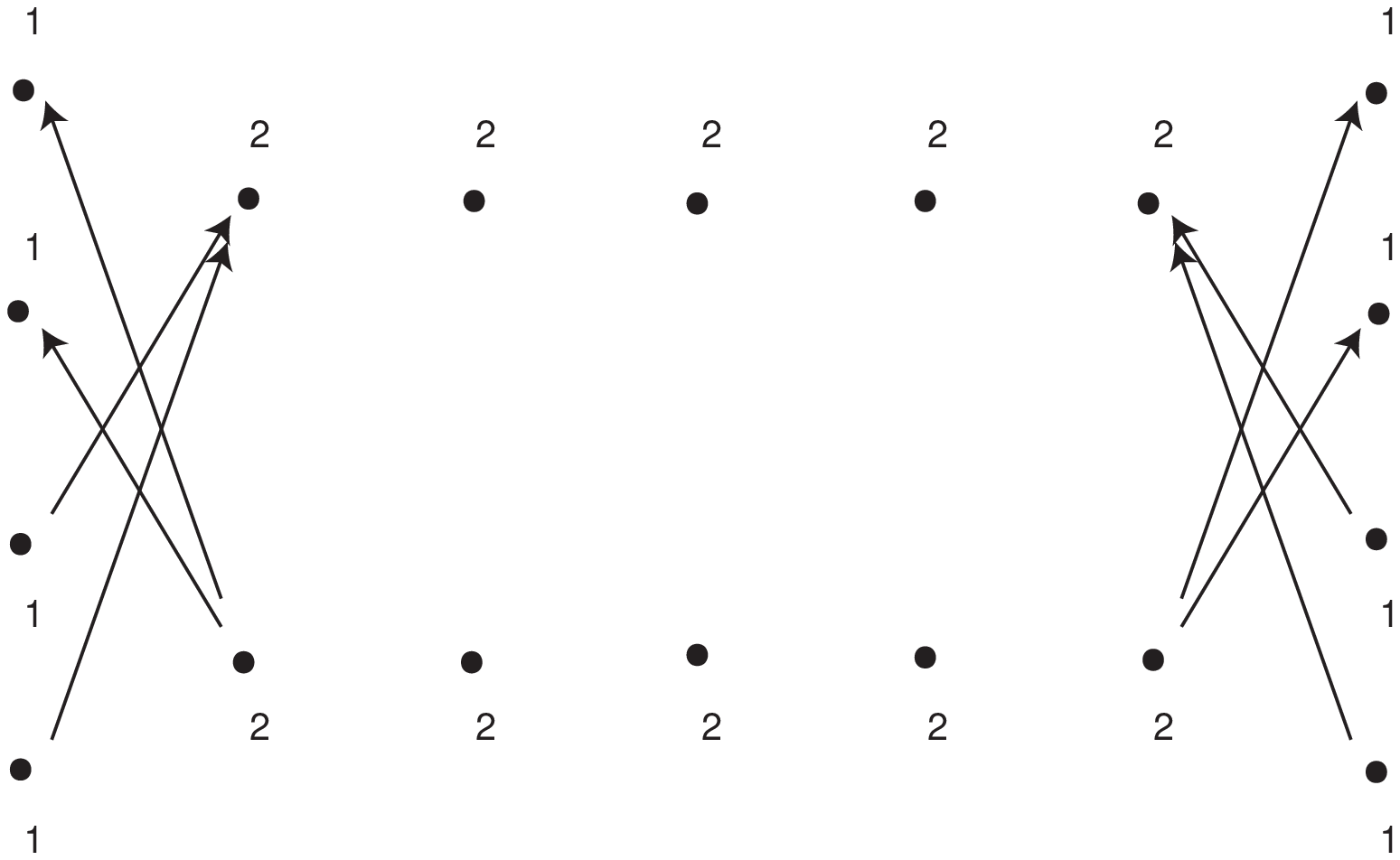}{8}{Connected nodes that can give rise to fractional branes at the $\BZ_n$ fixed point singularities.}
The charge that the magnetic flux carries in the node with higher Chern Simons level is canceled by the charge carried by the excitations of the fields represented by the arrows. The charge is split evenly between the other two nodes. This is natural from the point of view of taking a fractional brane from a $\BZ_{2n}$ singularity and projecting it onto a bound state of branes in a $D_n$ quiver that is obtained by folding the diagram (as in \cite{BL,BCD}).

We find this way that all the expected fractional branes at the singularity then have the same R-charge. This is a straightforward computation. We find also that for all  fractional branes a similar condition is satisfied to the
Martelli-Sparks \cite{MSparks} condition to get a four dimensional moduli space.

\begin{equation}
\sum_{i\in {f.b.}} M_i  k_i=0 \label{eq:fb}
\end{equation}
Here $k_i= \pm\kappa \dim(R_i)$, and we sum over the indices that correspond to a fractional brane. The $M_i$ are
the ranks of the corresponding gauge group. This condition is a consistency condition for being able to saturate the
total charge carried by the fluxes with the matter fields.

We should also remember that in M-theory compactifications on a circle, the fractional charge carried by a fractional brane can usually be modified by changing the Wilson lines of the enhanced gauge symmetry group at the ADE singularity \cite{Wdvd,Asp}.
In matrix quantum mechanical  models, this is done by changing the effective gauge coupling constants  \cite{BCD} . Here we find that the corresponding way of  changing the tension of the fractional branes is by changing the levels of the different Chern-Simons coefficients. To insure that the fractional brane survives, we need to keep constraints like those in (\ref{eq:fb}) for the corresponding brane. Notice that now, since the circle
bundle of the Hopf fibration over the base is twisted, we find that we are only allowed to have discrete values for these fractional brane tensions (these are related to the quantization of the Chern-Simons terms). These tensions need to be related to fluxes, rather than Wilson lines, because the bundle over the base that the branes see is different. The fractional branes need to have a twisted connection on the fiber to have different quantization conditions on a bundle than the D0-brane charges would provide.
This is how the field theory tells us that the Chern-Simons terms are generated by fluxes \cite{Aganagic}.
Also, the consistency requirement that the charges are cancelled can be reformulated in the geometry side by the usual statement that the total flux on the worldvolume of the brane should cancel \cite{FW,MMS}.

\section{Conclusion}

We have seen in this paper how the computation of the spectrum of magnetic monopole operators is useful in finding the topology of the moduli space of vacua of three dimensional field theory. In particular we have seen a very direct connection between these objects and points in the moduli space of vacua. This can be seen by solving classical equations of motion of a superconformal field theory on a cylinder, where we impose the equations that define BPS states at the classical level. These involve a slight
improvement of the equations that describe the classical moduli space, because gauge theory fluxes are quantized already at the classical level. It is indeed these quantization conditions that produce the different topologies when we change the level of the field theory.

The natural setup for these investigations was described in terms of matrix equations, with a natural action by complex conjugation. These systems of equations find their natural home in the realm of representation theory of associative
algebras. In the particular case we study, the theory of $C^*$ algebras is appropriate. This is just the name for algebras that have a natural conjugation that needs to be compatible with the representation. For the case of group actions, we found that the crossed product setups (essentially a very careful treatment of the method of images) where easier to analyze than just looking at the quiver algebra directly.

Of particular interest, we found that the detailed description of these configurations can be mapped directly to D-brane probes of the dual geometry, including brane fractionation at the singularities.
With these tools, we were able to argue that the tension of fractional branes at nonabelian orbifold singularities follows the same pattern than as expected from string theory considerations. We saw that this seems to require a nontrivial flux  for the potentials that couple to fractional branes, and that these tensions are directly correlated with the Chern-Simons levels of the different nodes of the quiver diagram  representing the theory.

It is natural to then ask what happens when we change the values of these fluxes and in particular how the moduli space is modified, as well as the patterns of brane fractionation. We have also not analyzed the setup in cases with discrete torsion. This is currently being investigated in \cite{BRomo}.

There are many other theories that are interesting to analyze and that do not arise from orbifolds of the basic ABJM theory. It would be interesting to see how these techniques can be applied in those cases, especially in situations where the fields of the theory do not have canonical dimensions.

We believe that there is still a lot of information to be obtained from studying BPS questions in three dimensional theories. However, we should not forget that the detailed study of the dynamics of these theories should produce additional information about the dynamics of M-theory and the locality of the theory in eleven dimensions. This is still mysterious from this setups.

\section*{Acknowledgements}

D. B. would like to thank I. Klebanov and J. Park for various discussions related to monopole operators.
M. R. would like to thank C. Beil, D. Morrison and P. Smith for discussions. D. B.  is supported in part by DOE under grant DE-FG02-91ER40618.

\end{document}